%% file: main.tex
\documentclass[twocolumn]{aastex62}
\usepackage{amsmath,amstext}
\usepackage[T1]{fontenc}
\usepackage{apjfonts} 
\usepackage[figure,figure*]{hypcap}
\usepackage{commath}
\usepackage{color}
\usepackage{mhchem, multirow, upgreek}
   \newcommand{\rah}{$^{\mbox{\scriptsize h}}$}
   \newcommand{\ram}{$^{\mbox{\scriptsize m}}$}
   \newcommand{\ras}{$^{\mbox{\scriptsize s}}$}
   \newcommand{\decd}{$^{\circ}$}
   \newcommand{\decm}{$'$}
   \newcommand{\decs}{$\farcs$}
   
   \newcommand{\beam}{$\theta_{\mbox{\scriptsize maj}}\times\theta_{\mbox{\scriptsize min}}$}

 	\newcommand{\eq}[1]{\begin{equation}#1\end{equation}}
	\newcommand{\Kelvin}{{\,\rm K}}
	\newcommand{\au}{{\,\rm au}}
	\newcommand{\braket}[1]{\left(#1\right)}
	\newcommand{\eqnref}[1]{Eq.~(\ref{#1})}
	\newcommand{\splitting}[1]{\begin{split}#1\end{split}}
	\newcommand{\GHz}{{\rm \, GHz}}
	\newcommand{\fref}[1]{\figref{#1}}
	\newcommand{\tref}[1]{Table~\ref{#1}}
	\newcommand{\dd}{{\rm d}}
	\newcommand{\cm}{\,{\rm cm}}
	\newcommand{\mm}{\, {\rm mm}}	
	\newcommand{\mum}{\,{\rm \upmu m}}
	\newcommand{\MJ}{{ \rm \, M_{\rm J}}}
	\newcommand{\pc}{{\,\rm pc}}
	\newcommand{\Msun}{{\, \rm M_\odot}}
		
	\newcommand{\sourcename}{IRAS04368+2557}
	
	\newcommand{\yr}{{\rm \,yr}}
\newcommand{\revision}[1]{\textcolor{red}{#1}}
\newcommand{\systemname}{L1527~IRS}
\newcommand{\kbeam}{0\farcs095$\times$0\farcs075;-82$^{\circ}$}
\newcommand{\qbeam}{0\farcs087$\times$0\farcs068;76$^{\circ}$}
\newcommand{\threebeam}{0\farcs155$\times$0\farcs068;-1.6$^{\circ}$}
\newcommand{\fourbeam}{0\farcs195$\times$0\farcs133;-5.2$^{\circ}$}
\newcommand{\sevenbeam}{0\farcs072$\times$0\farcs067;-11$^{\circ}$}
\newcommand{\subbeam}{0\farcs096$\times$0\farcs076;-82$^{\circ}$}

\received{December 29, 2019}
\revised{April 28, 2020}
\accepted{\today}
\submitjournal{ApJL}

\shorttitle{Substructure Formation in \systemname{}}
\shortauthors{Nakatani et al.}

\begin{document}

\title{Substructure Formation in a Protostellar Disk of \systemname{}}
\author[0000-0002-1803-0203]{Riouhei Nakatani}
\affiliation{RIKEN Cluster for Pioneering Research, 2-1 Hirosawa, Wako-shi, Saitama 351-0198, Japan}
\email{ryohei.nakatani@riken.jp}
\author[0000-0003-2300-2626]{Hauyu Baobab Liu}
\affiliation{Academia Sinica Institute of Astronomy and Astrophysics, P.O. Box 23-141, Taipei 10617, Taiwan} 
\author[0000-0002-9661-7958]{Satoshi Ohashi}
\author[0000-0001-7511-0034]{Yichen Zhang}
\affiliation{RIKEN Cluster for Pioneering Research, 2-1 Hirosawa, Wako-shi, Saitama 351-0198, Japan}
\author[0000-0002-7538-581X]{Tomoyuki Hanawa}
\affiliation{Center for Frontier Science, Chiba University, 1-33 Yayoi-cho, Inage-ku, Chiba, Chiba 263-8522, Japan}
\author[0000-0002-7570-5596]{Claire Chandler}
\affiliation{National Radio Astronomy Observatory P.O. Box 0,
1003 Lopezville Rd,
Socorro, NM 87801-0387, U.S. }
\author[0000-0002-0197-8751]{Yoko Oya}
\affiliation{Department of Physics, The University of Tokyo, 7-3-1, Hongo, Bunkyo-ku, Tokyo 113- 0033, Japan}
\author[0000-0002-3297-4497]{Nami Sakai}
\affiliation{RIKEN Cluster for Pioneering Research, 2-1 Hirosawa, Wako-shi, Saitama 351-0198, Japan}

\begin{abstract}
We analyze multi-frequency, high-resolution continuum data obtained by ALMA and JVLA to study detailed structure of the dust distribution in the infant disk of a Class~0/I source, \systemname. 
We find three clumps aligning in the north-south direction
in the $7 \mm$ radio continuum image. 
The three clumps remain even after subtracting free-free contamination,
which is estimated from the $1.3\cm$ continuum observations. 
The northern and southern clumps are located at a distance of $\sim 15\au$ from the central clump and are likely optically thick at $7\mm$ wavelength.  
The clumps have similar integrated intensities. %dust mass of $\gtrsim 0.4$--$0.5 \MJ$. 
The symmetric physical properties 
could be realized when a dust ring or spiral arms around the central protostar is projected to the plane of the sky.  
We demonstrates for the first time that such substructure may form even in the disk-forming stage, where the surrounding materials actively accrete toward a disk-protostar system. 
\end{abstract}


\section{Introduction}

	Since dust is the building blocks of planets, 	
	its spatial distribution is considered to directly link to the birthplaces of planets.
	Recent millimeter observations have revealed
	substructures in protoplanetary disks, such as rings, gaps, and spirals. 
	Substructures are found in all the 20 Class~II 
	targets in the Disk Substructures at High Angular Resolution Project (DSHARP) \citep[e.g.,][]{Andrews2018}.
	The results suggest that substructures are likely common among protoplanetary disks. 
	Investigating the origin is thus essential in the context of planet formation.

	Theoretical works have proposed several mechanisms that can be responsible for the substructures:
	torques due to massive planets \citep[e.g.,][]{Goldreich1980}, 
	secular gravitational instability \citep[e.g.,][]{Youdin2011}, 
	dust sintering \citep[e.g.,][]{Okuzumi2016}, etc., 
	but consensus has yet to be reached. %on the origin. 
	The verification requires observations to
	constrain when the substructure formation begins and how large dust have grown by that time. 
	To this end, 
	investigating dust distribution is necessary for younger sources with multiwavelength observations. 
	Recent continuum observations by the Atacama Large Millimeter/ submillimeter Array (ALMA)
	have detected ring structures in the disks around Class~I protostars \citep{Sheehan2017, Sheehan2018}. 
	However, it has not yet been known 
	if substructures can also form in even younger systems.

	The protostellar core L1527 is known to harbor a Class~0/I protostar, \sourcename.
	It has been reported 
	there exists a rotationally supported disk %inside the centrifugal barrier  
	\citep{Tobin2013, Sakai2014, Ohashi2014, Aso2017}.
	The disk size is $r \sim 80\au$ \citep{Oya2016, Oya2018}, 
	which is common among Class~0 sources \citep{Yen2015},
	while the relatively short distance of $137\pc$ \citep{Torres2007}
	makes it an optimal target to study substructure formation. 
	The envelope-disk system of \systemname{} is nearly edge-on with the disk slightly warped at $40\text{--}60\au$ \citep{Sakai2019}.
	In this study, the inner part of the warped disk ($r< 50\au$) is resolved by multiwavelength 
	observations with ALMA (Band 7, Band 4, Band 3) and JVLA (Q band, K band).

%-------------------------  Observations ---------------------
%\def \test#1{\ifx#1,\else 0 \expandafter\test\fi}
%\test{,}{test,}

\section{Observations} \label{sec:observation}

\subsection{ALMA Observations}	\label{sec:alma}

In the millimeter-wavelength range, we use ALMA
to observe the disk around the protostar, \sourcename. 
The observations have been carried out from 2015 to 2017.
The observation summary is presented in Tables~\ref{tab:obssum} 
and \ref{tab:qbandimage}.
The pointing and phase referencing centers were on 
R.A.$=$04\rah39\ram53\ras.870 (J2000), Decl.$=$ $+$26\decd03\decm09\decs6 (J2000).
We use Common Astronomy Software Applications \citep[CASA;][]{McMullin2007} package
for the calibration and analysis.
The reduction and calibration are done with CASA in a standard manner.

The proper motion of the target source is appreciable
\citep[$\mu_{\alpha}=$0.5 mas\,yr$^{-1} $, $\mu_{\delta}=$-19.5 mas\,yr$^{-1} $;][]{Loinard2002}.
To allow jointly imaging all data, and to compare the observations, 
we used the CASA task \textsc{fixplanet} to shift the target source to the expected coordinates on 01, Aug. 2017. 
(See \appref{app:observations} for the more details.)

\subsection{JVLA Observations}

We have retrieved the archival National Radio Astronomical Observatory (NRAO) Karl G. Jansky Very Lary Array (JVLA) observations towards \systemname{}.
The pointing and phase referencing centers were on R.A. $=$04\rah39\ram56\ras.600 (J2000), Decl.$=$ $+$26\decd03\decm06\decs00 (J2000).
% The observations have been carried out from 2011 to 2013.
They were carried out from 2011 to 2013, which were interleaved with Q band, K band, and C band observations in each epoch.
The observations utilized the standard continuum observing modes, which took full RR, RL, LR, and LL correlator products over a $\sim$2 GHz bandwidth coverage in 2011 using the 8-bit sampler, and over a $\sim$8 GHz bandwidth coverage using the 3-bit sampler in 2013.
The observations in 2011 adopted an 1-second integration time, while the observations in 2013 adopted an 3-seconds integration time.
Table \ref{tab:obssum} summarizes the details of these observations.
%The Q band and K band observations are carried out simultaneously. 
We give a summary of the JVLA observations in \tref{tab:obssum} and \tref{tab:qbandimage}.

% We manually calibrated the data following the standard calibration procedure, using CASA. 
%\baobab{
We manually calibrated the data following the standard calibration procedure, using the Common Astronomy Software Applications \citep[CASA;][]{McMullin2007} package (release 4.7.2). 
% The absolute fluxes of the flux calibrator 3C147 were referenced from the  Perley-Butler 2010 and  Perley-Butler 2013 flux standards \citep{Perley2013} for the observations taken before and after 2012, respectively.
%
% Since the target source is not bright enough to be eligible for gain phase self-calibration, 
% we have performed careful and extensive data flagging, in particular, for the A array configuration observations taken in the summer of 2011, to avoid distortion and blur in the final image due to phase error and dispersion. 
The proper motion of the target source is 
also taken into account. (See \secref{sec:alma}.)
We combined A and B configuration data of Q-band observation to obtain the map shown below. %in \secref{sec:results}.  
The map for each configuration data is also shown in \appref{app:jvlaobs}.
It also describes our data calibration in detail. Note that the data at $uv$ distances greater than 1000 $k\lambda$ was flagged for Q-band.

\section{Results}	\label{sec:results}

	\subsection{Observed Images}
\begin{figure*}[htbp]
\begin{center}\includegraphics[clip, width = \linewidth]{%figs/
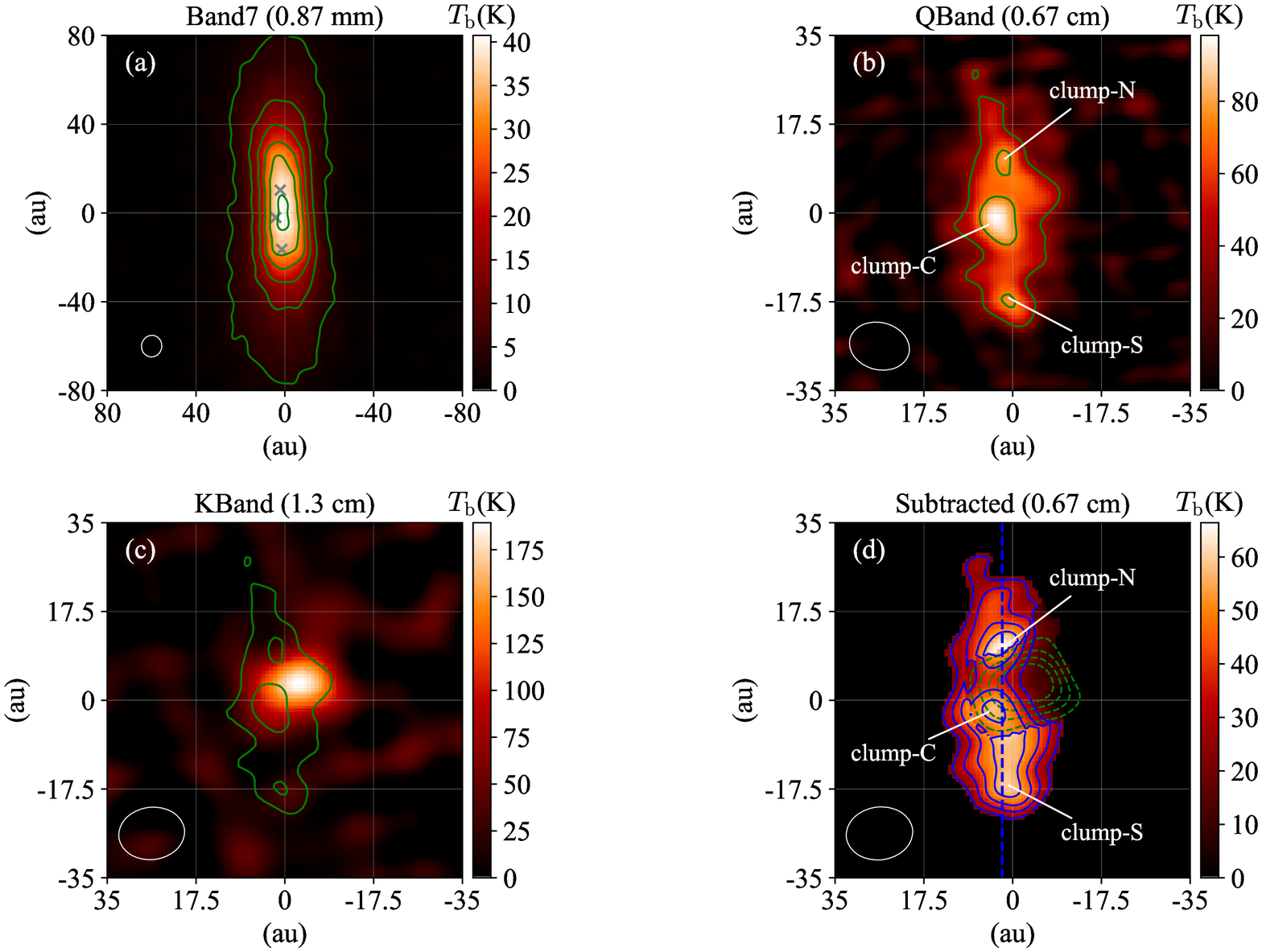}
\caption{
(a,b,c) Intensity maps for the observed data at ALMA Band 7, and JVLA Q- and K-bands, respectively.
The color map represents the brightness temperature. % $T_{\rm b}$. 
The horizontal and vertical axes indicate west-east and south-north directions.
The white ellipses at the bottom left indicate the beam sizes
(\beam; P.A.): \sevenbeam (Band~7), \qbeam (Q band), and \kbeam (K band).
The RMS noise level is $0.3\Kelvin$, $11\Kelvin$, and $21\Kelvin$
for Band~7, Q~band, and K~band, respectively.
In the ALMA Band~7 image,  
the green contours show the brightness temperature;
the first contour starts at $10\sigma$, and the interval is $23 \sigma$.
The clump locations are marked with the crosses. 
For the JVLA images, 
the green contours show the brightness temperature observed at Q band;
the first contour is at $3.5\sigma$ ($\sim 40\Kelvin$), and the interval is $3.0\sigma$. 
(d) the bottom right panel shows the Q band image where free-free contamination is subtracted. 
The beam size is \subbeam.
The solid blue contours in the subtracted map are plotted in the same manner as (b) and (c),
but with $1.5\sigma$ intervals and the RMS noise level in the smoothed Q band data, $\sigma'  \sim 7\Kelvin$.
The estimated free-free emission for Q band is indicated by the green dashed contours at 2, 3, 4, 5$\sigma'$.
The blue dashed line indicates the midplane,
which we define the north-south line %(i.e., P.A. of the line is $0^\circ$) 
passing through the peak position of clump-N in the subtracted image.
}
\label{fig:images}
\end{center}
\end{figure*}

	\figref{fig:images} shows the intensity maps 
	at ALMA Band~7, and JVLA~Q and K bands
	(ALMA Bands~3 and 4 images are shown in \appref{app:observations}). 
	The maps are scaled in units of $\au$
	for an assumed distance of $137\pc$ \citep{Torres2007}.
	The intensities are shown in terms of the brightness temperature, $T_{\rm b}$. 
	We have confirmed a smooth, flared edge-on disk structure \citep{Sakai2017} 
	in the Band 7 image. 
	The Band 7 image shows a flat structure in the inner region
	(nearly square shape evident from the $102\,\sigma$ contour) 
	despite the highest angular resolution,
	implying a hydrostatic disk. 
	Band 4 and 3 images show consistent structure with the Band 7 image.
	The Q-band data shows three aligned clumps from north to south 
	(hereafter clump-N, clump-C, and clump-S). 
%	The clump positions are marked by crosses in the Band 7 map. 
	K-band image shows a peaky structure
	unresolved by JVLA observations with A configuration.

	The clumps detected in the Q-band data can be indicative of substructure,
	but the noisy data necessarily requires further inspection
	for confirmation of the reality. 
	First of all, 
	the clumps are consistently resolved in independent multiepoch observations
	taken over a period of two years. 
	The combined A-configuration data (hereafter data~A) 
	and the three independent B-configuration data 
	individually show the southern clump 
	brighter than the northern clump (\fref{fig:qbandimage}). 
	The data~A
	and B2-configuration data (\tref{tab:qbandimage}), 
	which have the highest and the second highest angular resolutions in the north-south direction, respectively,
	show a clear gap between clumps-C and -S.
    The contrast between the gap and the peaks is
	as large as $\sim 4$--$5\sigma$ in data~A. 
%	\mycomment{How can I estimate the possibility of detecting artifacts at the same locations over independent multiepoch observations?}
	Second, we have observed high total peak S/N ratios ($\gtrsim 7$) 
	at the clumps-C and -S 
	for each of data~A and all the three B-configuration data. 
	With an assumption of the Gaussian noise statistics,
	probabilities of being a false positive is extremely low ($\lesssim 10^{-6}$). 
	We note that the level above the background intensity of the disk is $\sim 2\,\sigma$.
	Third, we find that all of the Q-band data can be well fit by triple 2D-Gaussians (see \fref{fig:2dgsfit} and \tref{tab:2dgsfit} for the fit results).  
	The peak positions of 2D-Gaussians are consistent, and 
	the deviation is quite small (see \tref{tab:2dgsfit}). 
	The peak distances are similarly consistent especially for that between clumps-C and -S. 
    Overall, the detection of the clumps and gap with a sufficient significance at consistent locations
    over multiepoch observations gives a high likelihood of being physical origin. 
    Detection of clump-N is relatively insignificant compared to clump-C and -S. 
    We discuss the geometry of the clumps in \secref{sec:origin}.

	Given that the detected emission is physical, 
	it is also necessary to assess possibilities that 
    a spurious clumpy structure is produced in a smooth disk by noise. 
	We have developed an infinitesimally-thin, smooth disk model 
	based on observed temperature profile for \systemname{} \citep{Tobin2013}
	and have performed synthetic observations 
	with integration times of $1$--$2$ hours 
	using the {\tt simobserve} task of CASA. 
	The surface density profile is given by a simple power-law as $\Sigma_0 (R/1\au)^{-1}$. 
	Since the mass of the inner disk is unknown, 
	we range $\Sigma_0$ in $10^3 {\rm \,g}\cm^{-2} \leq \Sigma_0 \leq 10^4{\rm \, g}\cm^{-2}$. 
	The inclination is set to $i = 5\deg$. 
	We also consider the case of $i = 10\deg$ to take account of disk flaring 
	in an approximate manner. 
	The total flux density of the model is $\sim 3$--$5 {\rm \,mJy}{\rm \, beam}^{-1}$.
	We have found that while the synthetic observations 
	with the VLA B-configuration resolve a smooth disk, 
	those with the A-configuration yield an apparent clumpy structure 
	extending in the north-south direction. 
	However,
	typically $50$--$70\%$ of the flux has been lost 
	in the A-configuration observations, while 
	the flux density is sufficiently recovered in the B-configuration observations
	even with an hour integration time.
	(We show an example of our synthetic observations for a smooth disk in \fref{fig:simobserve}.)
	Apparently, simultaneously reproducing 
	major features of the detected clumps
	in our actual observations with both of A- and B-configurations (\fref{fig:qbandimage})
	is likely difficult with a single smooth disk model,
	unless we may fine-tune parameters of the model and/or synthetic observations. 
	We have not found any of such fine-tuned parameters in our observations.
	A smooth disk model would not be capable of reproducing all of the observed features simultaneously.
	It appears that a smooth disk requires a series of coincidences to be observed as clumpy structure found in our Q-band data.

	We derive the spectral index $\alpha$ as
	\begin{gather}
		\alpha_{\lambda_1 - \lambda_2}  = \frac{\ln I_1 - \ln I_2}
						{\ln \nu_1 - \ln \nu_2},
%		\beta	    \equiv	\alpha - 2,
	\end{gather}
	where $\lambda_i$,
	$I_i$, and $\nu_i$ 
	denote the wavelength, intensity, and frequency at the $i$-th band:
	Band~7 ($0.9\mm$), Band~4 ($2\mm$), Band~3 ($3\mm$), Q~band ($7\mm$), and K~band ($1.3\cm$).
	To derive $\alpha$, we smooth the two images by the minimum beam that covers the beams of the original two data, and we use data with $3\sigma$ or higher detection.

    We find that the spectral indices between the ALMA data and the Q-band data
    show $\alpha \lesssim 2$ for the $< 20\au$ region,
    being consistent with optically-thick dust emission 
    (see Appendix\,\ref{sec:alphaindex} for more details).
\begin{figure}[htbp]
\begin{center}
\includegraphics[clip, width = \linewidth]{%figs/
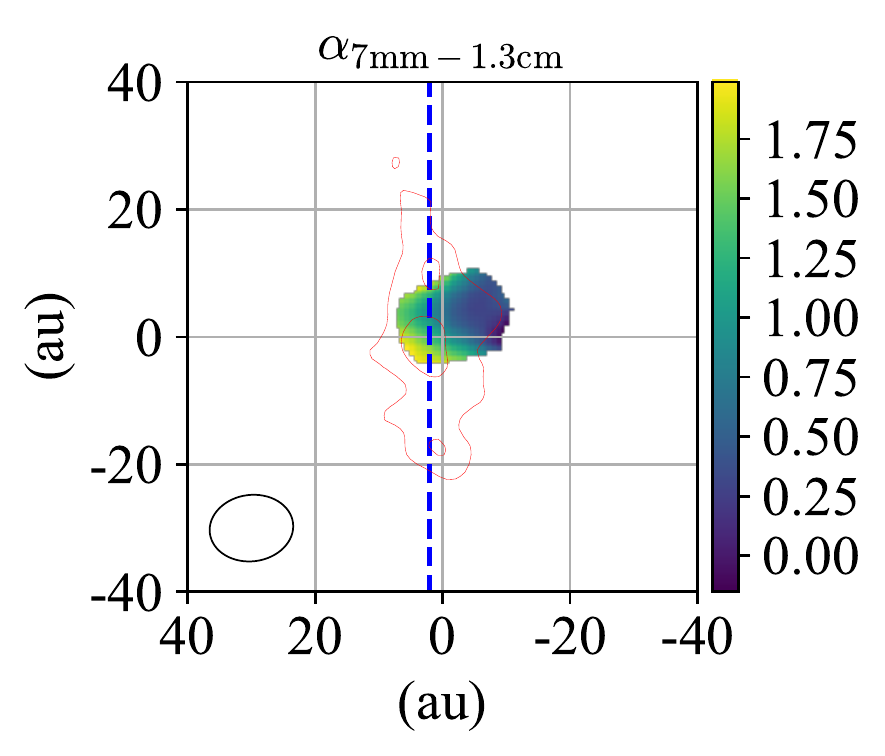}
\caption{
Map of the $\alpha$ index derived with the observation data at K and Q bands.
%The black region indicates less than $3\sigma$ detection at either of K band or Q band. 
The synthesized beam size is shown at the bottom left, and the size is \subbeam.
The red contours are shown in the same manner as \fref{fig:images}.
%The red and white contours represent the brightness temperatures at K and Q bands
%for the synthesized beamsize, respectively.
%The contours %are plotted with $40\Kelvin$ intervals from $(T_{\rm max} - 100\Kelvin)$ for each band, 
%indicate equi-temperature lines at 40\%, 60\%, and 80\% of the maximum for each band.
%where $T_{\rm max}$ is the maximum brightness temperature of the band data. 
}
\label{fig:betaKQ}
\end{center}
\end{figure}
    
	Typical spectral index  
	is $\alpha_{\rm 7\mm-1.3\cm} \sim 0$ above the midplane 
	between the lowest two bands (\fref{fig:betaKQ}). 
	It is consistent with that expected for an optically-thin free-free emission 
	from a thermal plasma protruding westward from the protostar. 
	The spectral index is close to $\sim 2$ at clump-C,
	where the emission is likely optically thick.
	The apparent offset between the K-band peak and clump-C
	may originate from the difference in the optical depth. 
	This offset has been consistently resolved in multiepoch observations
	with different array configurations (\fref{fig:qbandimage}), 
	and our calibrator fluxes are sufficiently accurate (\tref{tab:obssum}). 
	It suggests the offset to be physical.

	The C band observations have too poor angular resolution ($\sim0\farcs4$) to confirm the angular offset. In addition, we are not able to unambiguously determine the spectral index, because we cannot constrain the parameters of free-free emission (e.g., density, temperature, and emission measure) without degeneracy only with currently available data. Still, the estimated spectral index alpha between K and C band data ($\sim 0.4$--$0.7$) is consistent with optically-thin free-free emission at K band. Note that free-free emission can be optically thick at the C-band frequencies.
	The peak may indicate a current shock position or originate 
	from a highly opaque central region
	that attenuates the free-free emission from the other half of \ion{H}{2} region. 
	Note that the disk inclines by $i \sim 5 \deg$; 
	the western disk surface faces us \citep{Tobin2008,Tobin2010,Oya2015,Aso2017}.
%	If the disk is optically thick around the center, it may attenuate the free-free emission from 
%	the other half of the ionized regions.

	The $\alpha$ maps
	suggest that we observe optically-thick dust emission in the ALMA images
	and observe dust emission contaminated by free-free emission in the Q-band image. 
	To extract free-free contamination from the Q-band data, 
	we smooth the Q and K band images with the minimum beam 
	that covers both of the K- and Q-band beams
	to subtract the free-free contamination as
%	$0.095978arcsec-0.076077arcsec-97.071495deg$.
%	We approximately calculate pure thermal emission of dust as  
	\eq{
		I_{\rm 7\mm,dust} = I_{\rm 7\mm} - I_{\rm 1.3\cm} \braket{\frac{\nu_{\rm 7\mm}}{\nu_{\rm 1.3\cm}}}^{-0.1},
		\label{eq:freefree}
	}
	where the adopted spectral index, $-0.1$, 
	is typical for optically-thin free-free emission \citep[e.g.,][]{Anglada2018}.
	and is consistent with 
	the aforementioned measurements of $\alpha _ {7\mm\text{--}1.3\cm}$.
    The three clumps are evident in the subtracted intensity map
    regardless of a larger beam size than that of the original Q-band image (\fref{fig:images}d).
	Both of the clump-N and -S are located 
	at a distance of $\sim 15\au$ from clump-C. 
	The brightness temperatures are similar ($\simeq 60 \Kelvin$) for all the clumps. 
	The clump-C temperature may be underestimated
	because \eqnref{eq:freefree} likely overestimates 
	the contamination at the clump-C position (cf.~\fref{fig:betaKQ}).

\section{Discussions}	\label{sec:discussion}
%	We have found substructure in the disk of a Class~0/I source, \systemname{}, 
%	with the high sensitivity and high spatial resolution observations by JVLA. 
%	The clump locations are symmetric with respect to the center,
%	and the projected mass is likely similar. 
%	In this section, we discuss plausible modality and the possible origin of the substructure.
    The clumpy structure shown in the Q-band image 
    can be indicative of substructure in \systemname{}. 
    In this section, the physical properties of the clumps are examiend
    to consider the actual clump geometry and possible origins. 
    We first present our measurements of the optical thickness (\secref{sec:optdep})
    and the mass (\secref{sec:masses}),
    then discuss the geometry and origins (\secref{sec:origin}).

    \subsection{Optical Thickness}   \label{sec:optdep}
\begin{figure}[htbp]
	\begin{center}
	\includegraphics[clip, width = \linewidth]{%figs/
	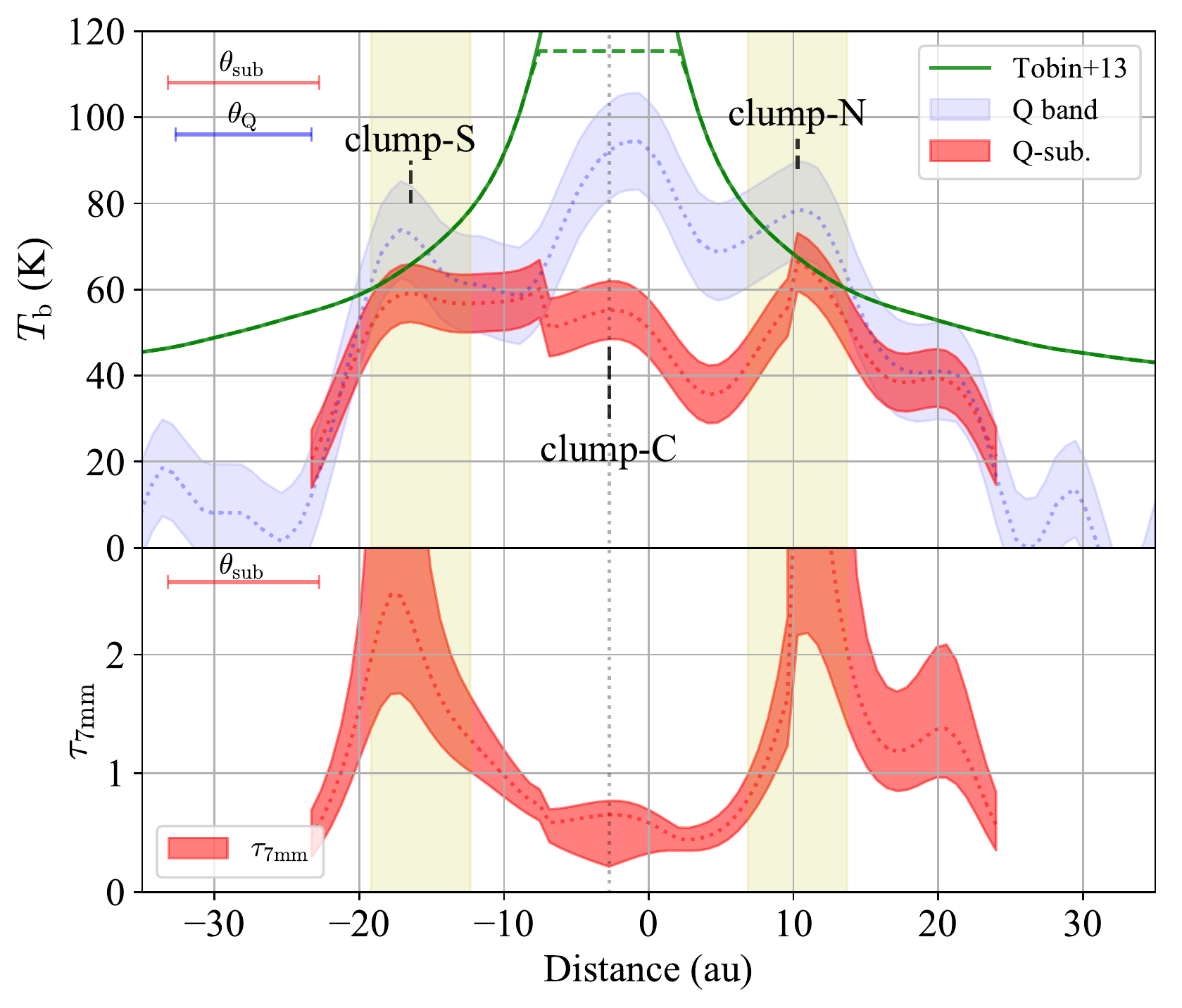}
	\caption{(top) distribution of the brightness temperature along the midplane.
	The red and blue dotted lines show those of the subtracted Q band and the original Q band, respectively.
	The filled color regions represent $\pm 1\,\sigma$ ranges.  
	The temperature model of \cite{Tobin2013}  
	and the modified model are indicated by the solid and dashed green lines, respectively. 
	We also show the sliced beam sizes 
	for the subtracted Q band data $(\theta_{\rm sub})$ and the original Q band data $(\theta_{\rm Q})$
	at the top left. 
	The yellow shaded regions indicate the radial positions for the \ce{CO2} snow line ($60$--$80\Kelvin$). 
	(bottom) the red shaded region shows $\tau_{\text{\scriptsize 7\,mm}}$ 
	resulting from $T_{\rm b}$ of the subtracted data and the model temperatures.
	%The estimated Toomre $Q$ parameter is shown by the black dashed line (see \secref{sec:discussion}).
	%The grey shaded area represents the upper and lower limits.
	}
	\label{fig:taunu}
	\end{center}
	\end{figure}

%	We estimate the optical thickness of the clumps
%	to investigate the origins of the substructure. 
    We adopt the opacity model of \cite{Birnstiel2018} in our discussions. 
	Given that we have not detected evidence of dust growth in the system
	and it is also the case for Class~II disks \citep[e.g.,][]{Kataoka2016,Liu2019}, 
	we assume $a_{\rm max} \lesssim 100 \mum$ in the following discussions.  
	\footnote{	Scattering opacity is negligible for $a_{\rm max} \lesssim 100\mum$, 
	while it dominates over absorption opacity for $a_{\rm max} \gtrsim 400 \mum$. 
	}
%	(See also \secref{sec:masses} for possible uncertainties of the adopted opacity.)
	Using the temperature model of \cite{Tobin2013},
%	based on infrared- and submillimeter-wavelength observations with a spatial resolution of $\sim 40\au$,
	we estimate $\tau_{\text{\scriptsize 7\,mm}}$ along the midplane as
	\eq{
		\tau_{\text{\scriptsize 7\,mm}} = - \ln \braket{1 - \frac{T_{\rm b,7\mm}}{T_{\rm model}}},
	}
	where $T_{\rm model}$ is the model temperature. 
	The model temperature goes extremely high around clump-C,
	which can lead to a significant underestimation of $\tau_{7\mm}$.
	To approximately take account of a possible distance between clump-C and the protostar,	
	we also use a modified temperature distribution, $T_{\rm model}'$,
%$	to derive $\tau_{\text{\scriptsize 7\,mm}}$.
%	We modify $T_{\rm model}$ 
	by truncating $T_{\rm model}$ from the top within $5\au$, 
	which corresponds to the gaussian-fit scale of clump-C. 
	The resulting $\tau_{\text{\scriptsize 7\,mm}}$ is shown in \fref{fig:taunu}. 
	The regions around clump-N and -S appear to be optically thick ($\tau_{\text{\scriptsize 7\,mm}} > 1$);
	the gap regions ($d\lesssim 15\au$) are marginally optically thick $\tau_{\text{\scriptsize 7\,mm}} \sim 1$,
	and the outer ($d \gtrsim 20 \au$) regions are optically thin $\tau_{\text{\scriptsize 7\,mm}} < 1$. 
%    Since the disk is likely optically thin at $d\sim 40\au$ for $7\,\mm$ wavelength,
	
	We have found 
	$\alpha$ values of 
	$\alpha_\text{3mm-7mm}$ ($\simeq 2.5$--$2.8$) at $d = 30\au$ 
	and $\alpha_\text{0.9mm-3mm}$ ($\simeq 2$--$2.5$) at $d=40 \au$
	(cf.~\figref{fig:alphamaps}).
	The $\alpha$ values are 
	much smaller than the typical interstellar value \citep[$\alpha\approx3.8$;][]{Draine2006}.
	%could reflect dust property.
	%The small values of $\alpha_\text{3mm-7mm}$ ($\simeq 2.5$--$2.8$) at $d = 30\au$  and $\alpha_\text{0.9mm-3mm}$ ($\simeq 2$--$2.5$) at $d=40 \au$
	It could suggest dust growth from the interstellar dust
	at the outer region of \systemname{}'s inner disk ($\gtrsim 30$--$40\au$).

\subsection{Clump Masses}    \label{sec:masses}
	We estimate the dust mass at clump-N and -S as
	\eq{
	\splitting{
		M_{\rm dust} =% -\frac{1}{ \kappa_{7\mm}} \int _A \dd A \, \ln\braket{  1 - \frac{I_{7\mm}} {B_{7\mm} (T_{\rm model})}}, \label{eq:estimatedmass}
		 & 0.59  \MJ \, \braket{\frac{\kappa_{7\mm}}{0.02\cm^2{\,\rm g}^{-1}}}^{-1} \\ %\tilde{ \kappa}^{-1}
		 & \times  \int _A \frac{\dd A}{(10\au)^2} \, \ln\braket{  1 - \frac{I_{7\mm}} {B_{7\mm} (T_{\rm model})}}^{-1}, \label{eq:estimatedmass}
		 }
	}	
	where $A $ is an integration area, 
	$B_\nu$ is Planck function, $\kappa_{7\mm}$
	is absorption opacity at Q~band.
	Note that the absorption opacity is 
	$\approx 0.02 {\rm \, cm^2 \, g^{-1}}$ for $a_{\rm max}  \lesssim 100 \mum$ \citep{Birnstiel2018}. 
	Since the clumps are not resolved, we set $A$ to the synthesized beam size
	whose center is located at the peak of each clump. 
	Adopting $\kappa_{7\mm} = 0.02 \cm^2 {\rm \,g}^{-1}$, 
	we estimate the lower limits of the clump mass to be 
	$0.40 \MJ$ for clump-N and $0.48 \MJ$ for clump-S.
	The lower limit of the total dust mass (with $3\sigma$ or higher detection) is 
	$2.7 \MJ$. %$2.7 \tilde{\kappa}^{-1}\MJ$. 

	The total dust mass of %$2.7 \MJ$ 
	corresponds to the total disk mass of $M_{\rm disk} \sim 0.26\Msun$
	for dust-to-gas mass ratios of 1\%,
	which implies that 
	that disk mass accounts for a large fraction ($\sim$50--100\%) 
	of the system's mass estimated from the rotation curve of $\sim 1\mm$ observations in prior studies 
	\citep[$\approx0.2$--$0.46\Msun$;][]{Sakai2014, Ohashi2014, Aso2017}.
	The inner disk has turned out to be likely optically thick at the band in this study,
	and thus its contribution to the system mass could have been underestimated. 
	
\if0
	Our estimated disk mass implies a high disk-to-star mass ratio of the order of unity,
	which is consistent with the recent observations of Class~0 sources that show %that Class~0 sources have
	a nearly unity disk-to-star mass ratio \citep{Li2017}
	\revision{\citep{Yen2017}}
	and similar disk mass functions to the stellar initial mass function \citep{Galvan2018}. 
	A high disk-to-star mass ratio is also seen in some recent MHD simulations \citep[e.g.,][]{Zhao2018}.
\fi

	An important note is that our estimated mass suffers from
	huge uncertainties mainly originating from poor understanding of dust 
	opacity at cm-wavelengths. 
	%The adopted model temperature has been developed based on infrared 
	%and sub-millimeter observations with $\sim 40\au$ resolutions. 
	%Besides, 
	%we hardly understand what opacity is preferable to measure mass at cm-wavelengths. 
	The adopted dust opacity may be low, leading us 
	to obtain a high mass for the clumps and disk. 
	% \eqnref{eq:estimatedmass} holds for absorption-dominant cases. 
	The opacity model of \cite{Woitke2016} predicts about an order of magnitude larger absorption opacity than the adopted one in this study. 
	(Scattering dominates at cm-wavelengths in their model, however.)
	In addition, the dust-to-gas mass ratio is also uncertain. 
	We stress that the estimated mass works as no more than a reference. 
	Future observations and models are indispensable 
	%to draw more meaningful conclusions on grain growth, opacity, temperature profile,
	%and mass in the region of interest. 
	for an accurate mass measurement. 
	Nevertheless, we show our measurement results, 
	expecting it to be a reference for future studies.

\subsection{Clump Geometry and Origins}	\label{sec:origin}
%\section{Discussion}	\label{sec:discussion}
	
%	We have found substructure in the disk of a Class~0/I source, \systemname{}, 
%	with the high sensitivity and high spatial resolution observations by JVLA. 
%	The clump locations are symmetric with respect to the center,
%	and the projected mass is likely similar. 
%	In this section, we discuss plausible modality and the possible origin of the substructure.

    The previous 7\,mm VLA observations of \cite{Loinard2002} have detected the disk as an elongated component
    with the data sets obtained on 1996 December 31 and 2002 April 10.
%	The study also found an additional unresolved blob 
%	to the east of the disk, but it has not been confirmed by other observations including ours.
%	The derived disk radius in \cite{Loinard2002} ($\sim 20\au$)
	%at which the intensity is half the maximum of the disk emission, 
%	is well consistent with our derived distances of the clumps ($\sim 15\au$)
%	which are defined as the distance between the peak positions of clump-C and clump-N/-S.
%	Given that the detected emission of \cite{Loinard2002} is real for the disk,
	The extension of the disk is largely similar to 
	those of clump-N and -S,
	\footnote{
    \cite{Loinard2002} also found an additional unresolved blob 
    to the east of the disk, which has not been confirmed by other observations including ours.
	}
	which implies that 
	the projected distances remain the same %within the beam size accuracy
	for about twenty years.
	The symmetries in the locations, mass, and time of clump-N and -S may indicate consistency with an axisymmetric geometry
	rather than individual chunks. 	
	Note that clump-N is not so bright as clump-S or clump-C (\fref{fig:qbandimage} and \fref{fig:2dgsfit}). 
	Possibilities of being individual chunks 
	are not ruled out.

	If the Q-band data shows an axisymmetric structure,
	a dust ring and symmetric spiral arms are plausible candidates.
	We derive Toomre $Q$ parameter \citep{Toomre1964} by approximately estimating surface density 
	as $\bar{\Sigma}  \approx 100 \, M_{\rm dust}(\leq d)/\pi d^2$ (cf.~\eqnref{eq:estimatedmass}).
%	The estimated Toomre $Q$ parameter (cf. the lower panel in \fref{fig:taunu}) 
%	suggests that the disk might be gravitationally unstable at the clump positions.	
%	Therefore, the spiral arms would be possible for the clump modality. 
    The $Q$-values are much lower than unity for $<15\au$
%    to be strongly unstable. 
    such that the disk causes significant fragmentation if the equation of state (EOS) is approximated to be isothermal.
    Emission would not be observed as a disk for a long period of time if this is the case. 
    It could indicate errors in our mass measurements or that the substructure is actually individual clumps. 
%    We might not be able to draw meaningful conclusions about the disk stability only
%    with currently available data. 
    In any case, 
    there are uncertainties in our mass measurements,
    and thus
    disk stability would be a matter of discussion after an accurate measurement is conducted with 
    high S/N data obtained by future observations.
    Besides, validating isothermal EOS would not be trivial in strongly accreting systems as \systemname{}. 
	Note that although the disk-to-star mass ratio is estimated to be high, 
	such disks do not necessarily fragment \citep{Kratter2010}. 
    %\revision{Radiation hydrodynamics simulations with self-consistent thermochemistry would be needed to understand gravitational stability of Class~0 disks.}

%    The derived $Q$ also suffers uncertainties for the same reasons discussed in \secref{sec:masses}. 
	The observed clumps are consistent with a projected dust ring
	if the disk is stable or marginally gravitationally unstable,
%	Given that the clumps represent a projected dust ring, 
	leaving the origins as an open question. 
	Dust sintering is possible to form a dust ring \citep{Okuzumi2016, Okuzumi2019}. 
%	Grains with $\sim 100\mum $ can accumulate at just outside of 
%	\ce{CO2} snow line ($60$--$80\Kelvin$) owing to both 
%	the sintering effects and the low stickiness of CO-mantled dust \citep{Okuzumi2019}.
	Interestingly, our small $\alpha$ values ($\sim 2.5$)
	at $d \gtrsim 30$--$40\au$
	and nearly identical locations of the clumps to \ce{CO2} snow line (\fref{fig:taunu})
	are compatible with this scenario. 
	Secular gravitational instability \citep[SGI;][]{Youdin2011, Takahashi2014} is another possible 
	explanation for a dust ring. 
	Since \systemname{} is an infall-dominant source, the age would be $\sim 10^4$--$10^5\yr$,
	which is comparable to the typical growth timescale of SGI.  % need to confirm it RN.
%	SGI can work in the disk of \systemname{} in this sense.  %if the diffusivity of dust due to gas turbulence is weak. 

	For further investigation, 
	we need gas kinematics observations with $<30\au$ resolution
	and continuum observations at longer wavelengths with high angular resolutions 
	to examine dust growth in the gap regions.

\section{Summary}	\label{sec:summary}
	We analyze high-resolution dust continuum data of \systemname{} obtained 
	by ALMA (Bands 3, 4, and 7) and JVLA (Q, K, and C bands).
	We have found three clumps aligning north to south in the Q-band data.
	The clumps are consistently detected in independent multiepoch observations with different array configurations
	over a period of two years. 
	We have concluded that the clumpy structure is physical origin
	and are indicative of substructure in \systemname.
%	The mass has also been estimated for the clumps and found to  be similar with $\gtrsim 0.4$--$0.5\MJ$
%	while there is a large uncertainty originating from
%	unknown properties of temperature and grain growth. %$\gtrsim 0.02$--$ 0.03 \MJ$. 	
	The north and south clumps are symmetrically located at a distance of $\sim 15\au$
	with respect to the central clump
	and are likely optically thick. 
	The integrated intensities are also similar. 
	The symmetric characters 
	propose a symmetric geometry such as 
	a dust ring and symmetric spiral arms.
	However, considering less brightness of the northern clump compared to the southern clump, 
	possibilities of being independent clumps are not rules out. 
	The origins of the substructure remain unclear. 
	Observing gas kinematics and dust continuum at lower-frequency bands is essential to address the issue. 
	Most importantly, 
	our results demonstrate that substructure formation can occur at the earliest stage of protostar-disk system formation.

\par

\acknowledgments %

We thank Satoshi Okuzumi, Hiroshi Kobayashi, Sanemichi Takahashi, 
and Hideko Nomura for fruitful discussions. 
We are also grateful to the anonymous referee for giving us practical and insightful comments,
which have greatly improved this manuscript. 
R.N. is supported by the Special Postdoctoral Researchers (SPDR) Program at RIKEN
and by Grant-in-Aid for Research Activity Start-up (19K23469). 
H.B.L. is supported by the Ministry of Science and Technology (MoST) of Taiwan (Grant Nos.~108-2112-M-001-002-MY3).
S.O. is supported by SPDR Program and JSPS KAKENHI Grant No.~18K13595. 
Y.Z. is supported by SPDR Program and JSPS KAKENHI Grant No.~JP19K14774.
N.S. is supported by Grant-in-Aid for Scientific Research (S) Grant No.~18H05222. 
The National Radio Astronomy Observatory is a facility of the National Science Foundation operated under cooperative agreement by Associated Universities, Inc.
This paper makes use of the following ALMA data: ADS/JAO.ALMA\#2013.1.00858.S, ADS/JAO.ALMA\#2016.A.00011.S, ADS/JAO.ALMA\#2016.1.01203.S, ADS/JAO.ALMA\#2017.1.00509.S. ALMA is a partnership of ESO (representing its member states), NSF (USA) and NINS (Japan), together with NRC (Canada), MOST and ASIAA (Taiwan), and KASI (Republic of Korea), in cooperation with the Republic of Chile. The Joint ALMA Observatory is operated by ESO, AUI/NRAO and NAOJ.
%\mycomment{To be updated.}

\facility{JVLA, ALMA}
\software{ CASA \citep{McMullin2007}}

\bibliographystyle{yahapj}
\bibliography{references}

%\begin{appendix}
\appendix

\section{Observation Summaries}	\label{app:observations}
\subsection{ALMA}

%Details of these two epochs of observations are summarized in Table \ref{tab:obs}.
%Combining the two epochs of observations can provide a $\sim$0\farcs2 synthesized beam and a 4$''$ largest recoverable angular scale.

The two epochs of observations shared an identical spectral setup, which provided five spectral windows at [146.965, 150.494, 150.432, 138.175, 138.346] GHz central frequencies with [58.594, 58.594, 58.594, 58.594 937.500] MHz frequency widths and [15.259, 30.518, 30.518, 15.259, 488.281] kHz frequency channel widths.

The observations of Epoch~1 were calibrated using  the Common Astronomy Software Applications \citep[CASA;][]{McMullin2007} package (release 4.7.0) and with Pipeline-Cycle4-R2-B.
The gain calibrator selected for Epoch~2 was faint, such that the standard pipeline failed to produce usable calibrated visibilities. 
We manually calibrated the observations of Epoch~2 using the 5.4.0 release of CASA.
After implementing the antenna position corrections, watervapor radiometer radiometer solution, and system temperature table, we performed the passband calibrations.
To yield reasonably high signal-to-noise (S/N) ratios when deriving the gain phase solutions, we first solved the phase offsets among spectral windows using the passband calibration scan.
After applying the passband and phase offsets solution, we then derived the gain phase and amplitude solutions by combining all spectral windows.
We derived the absolute flux scaling factors for the individual of the five spectral windows by querying the fluxes of the calibrator J0510+1800 from the calibrator grid monitoring survey of ALMA.
Finally, we performed three iterations of gain phase self-calibration using the spectral window which has the broadest frequency width, and then applied the solutions to all five spectral windows.

\begin{figure*}[htbp]
\begin{center}
\includegraphics[clip, width = \linewidth]{%figs/
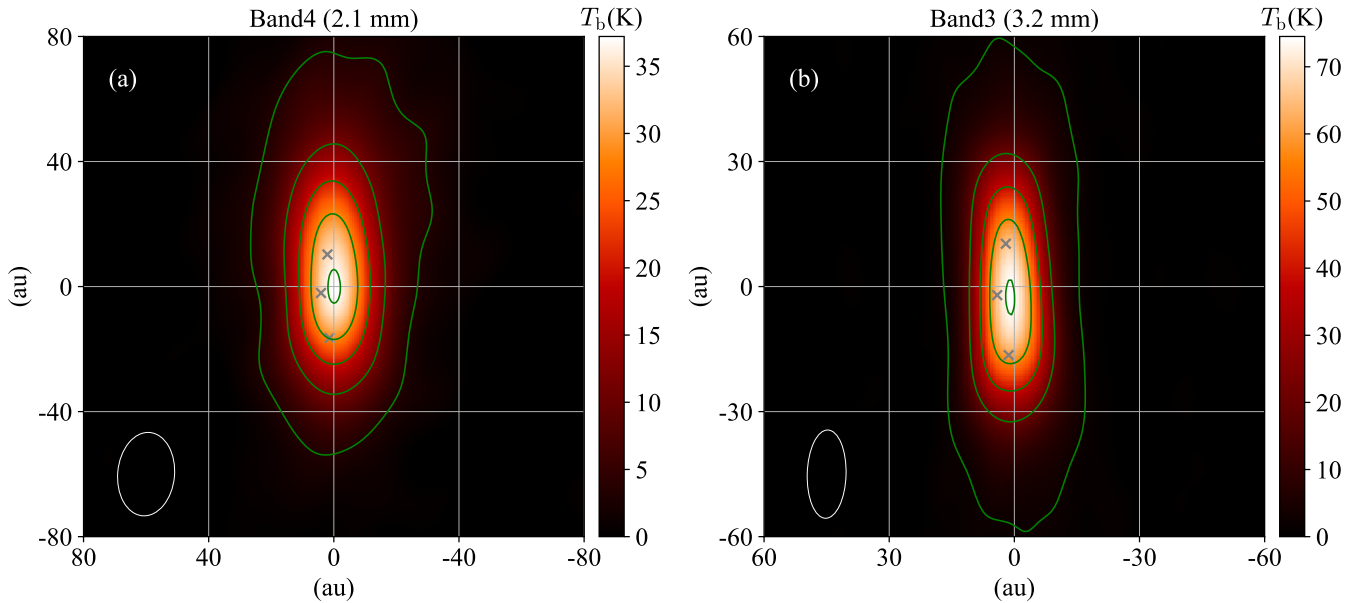}
\caption{
Brightness temperature maps for ALMA Band 4 and Band 3 observations. 
The beam sizes are \fourbeam and \threebeam,
and the RMS noises are $0.3\Kelvin$ for both of Bands~4 and 3, respectively.
The green contours are shown in the same manner as the ALMA Band 7 image in \fref{fig:images},
but the interval is $25\sigma$ and $65\sigma$ for the Bands 4 and 3 images, respectively.}
\label{fig:almaimage43}
\end{center}
\end{figure*}

\subsection{JVLA}	\label{app:jvlaobs}
% The observations utilized the standard 3-bit sampler continuum observing mode, which took full RR, RL, LR, and LL correlator products over a $\sim$2 GHz bandwidth coverage in 2011, and over a $\sim$8 GHz bandwidth coverage in 2013.
% Table \ref{tab:obssum} summarizes the details of these observations.
%\baobab{

We split the Q, K, and C band data from individual epochs of observations before calibrating them separately.
Following the standard procedure, we first implemented the corrections for antenna positions, the weather information, the gain-elevation curve, and the opacity model to all data.
Afterwards, we performed per-integration gain phase calibration for the absolute flux and passband calibrators, and then bootstrapped the delay fitting and passband calibration by referencing the absolute flux from quasar 3C147.
We adopted the Perley-Butler 2010 and  Perley-Butler 2013 flux standards \citep{Perley2013} for the observations taken before and after 2012, respectively.
After applying the delay and passband solutions, we derived the per-integration  gain phase solutions for all calibrators, and derived the per-scan complex gain solutions for the gain calibrator.
We applied the per-integration gain phase solution when deriving the per-scan gain amplitude solutions for all calibrators, and then derived the absolute flux calibration factors based on the per-scan gain amplitude solutions by referencing to the aforementioned 3C147 flux standards.
Finally, we applied the delay, passband, per-scan gain amplitude solution, per-scan gain phase solution, and the absolute flux calibration factors to the observations on our target source.

Owing to that the target source is not bright enough to be eligible for gain phase self-calibration, careful and extensive data flagging was performed, in particular, for the  A array configuration observations taken in the summer of 2011, to ensure that the final images were not significantly distorted or un-sharpened due to phase error and dispersion.
More specifically, we flagged the Q band (44-46 GHz) data which were taken in 2011 at below 35$^{\circ}$ elevation, and flagged the Q band data taken on 2011 August 06 which were taken with the $>$1000 $k\lambda$ projected baselines.
As a consequence, all 44-46 GHz data taken on 2011 July 21 were flagged.

\begin{figure*}
    \hspace{-0.2cm}
    \begin{tabular}{ p{5.7cm} p{5.7cm} p{5.7cm} }
        \includegraphics[width=5.7cm]{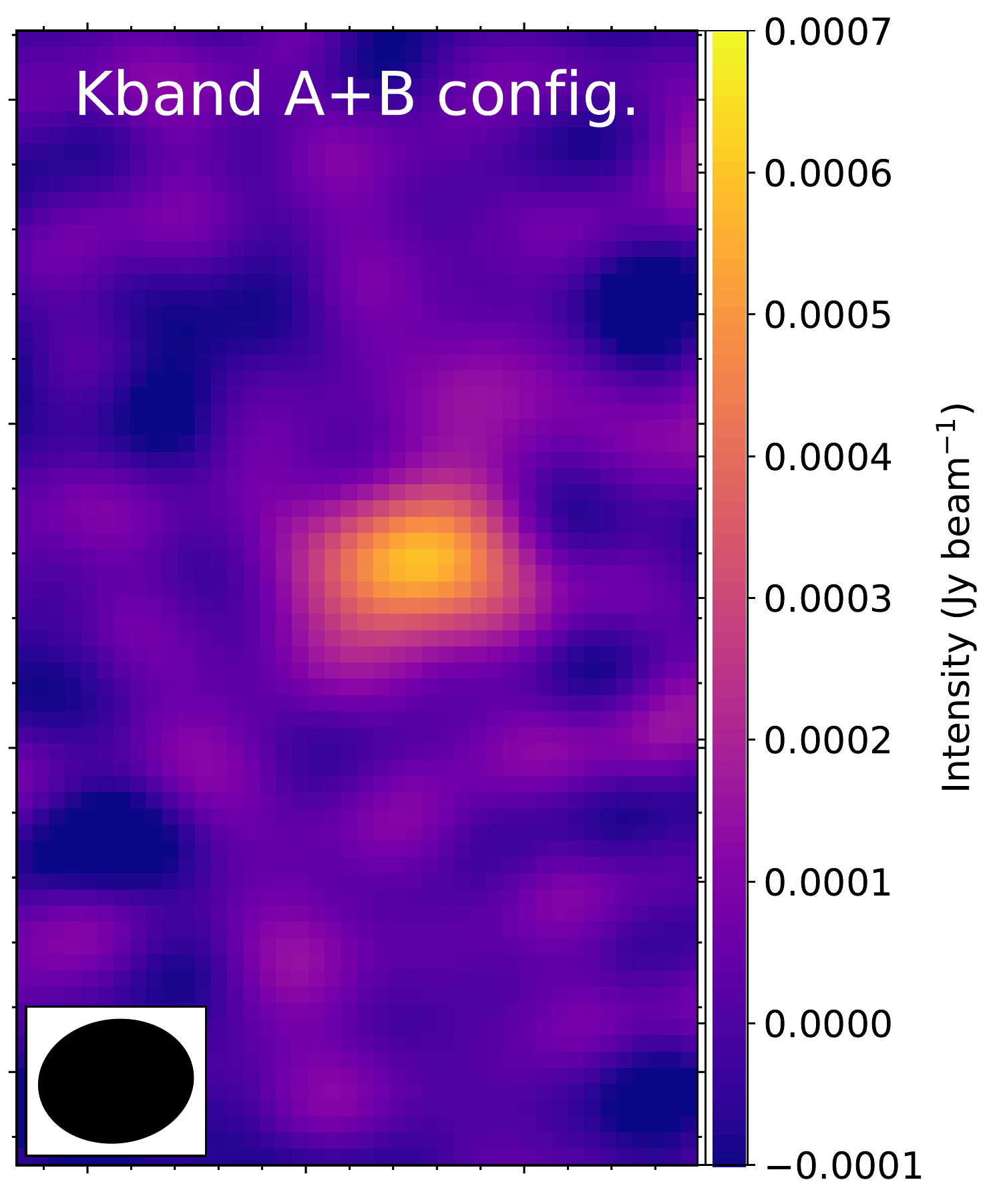} &  
        \includegraphics[width=5.7cm]{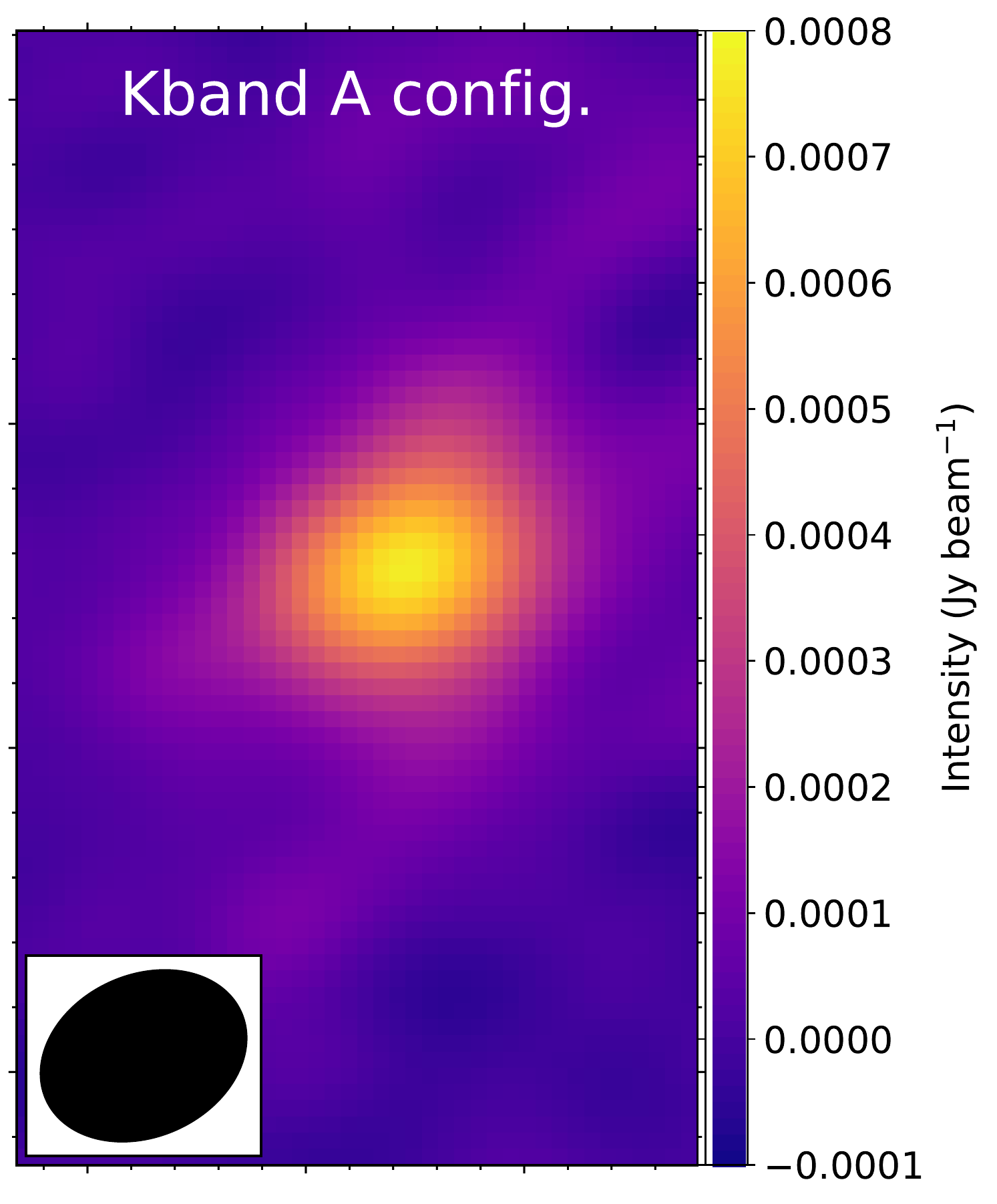}   & 
        \includegraphics[width=5.7cm]{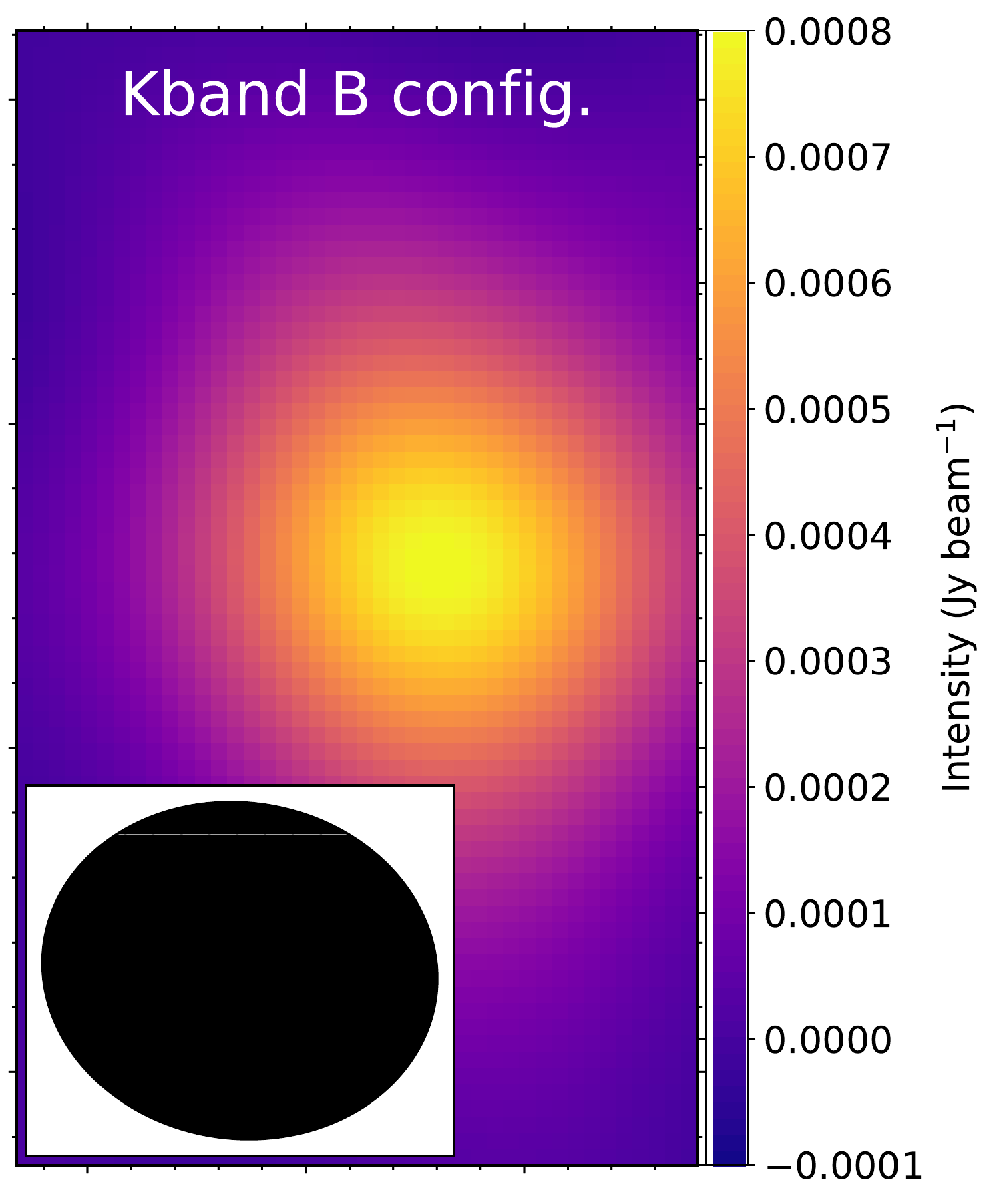}  \\
        \includegraphics[width=5.7cm]{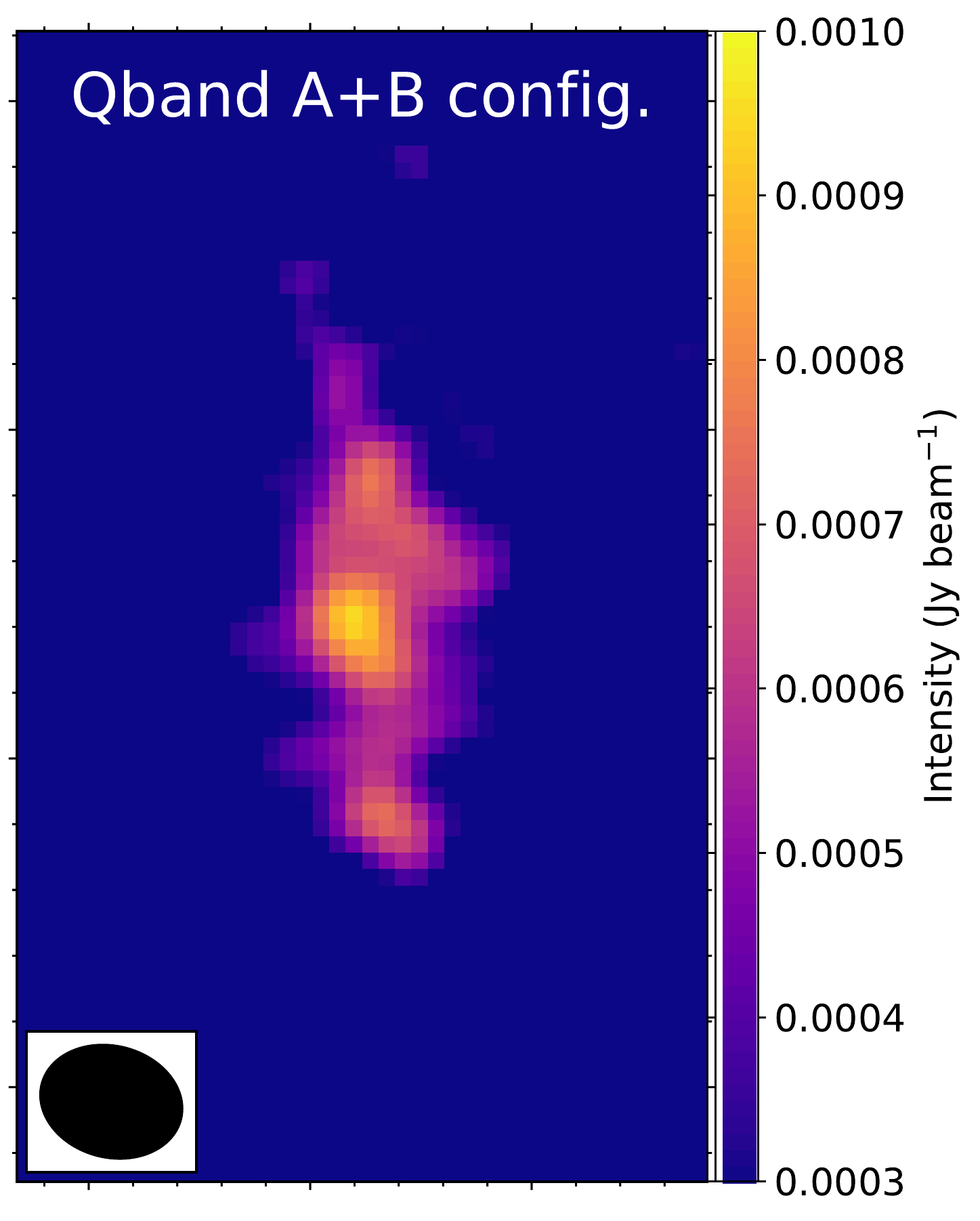} &  
        \includegraphics[width=5.7cm]{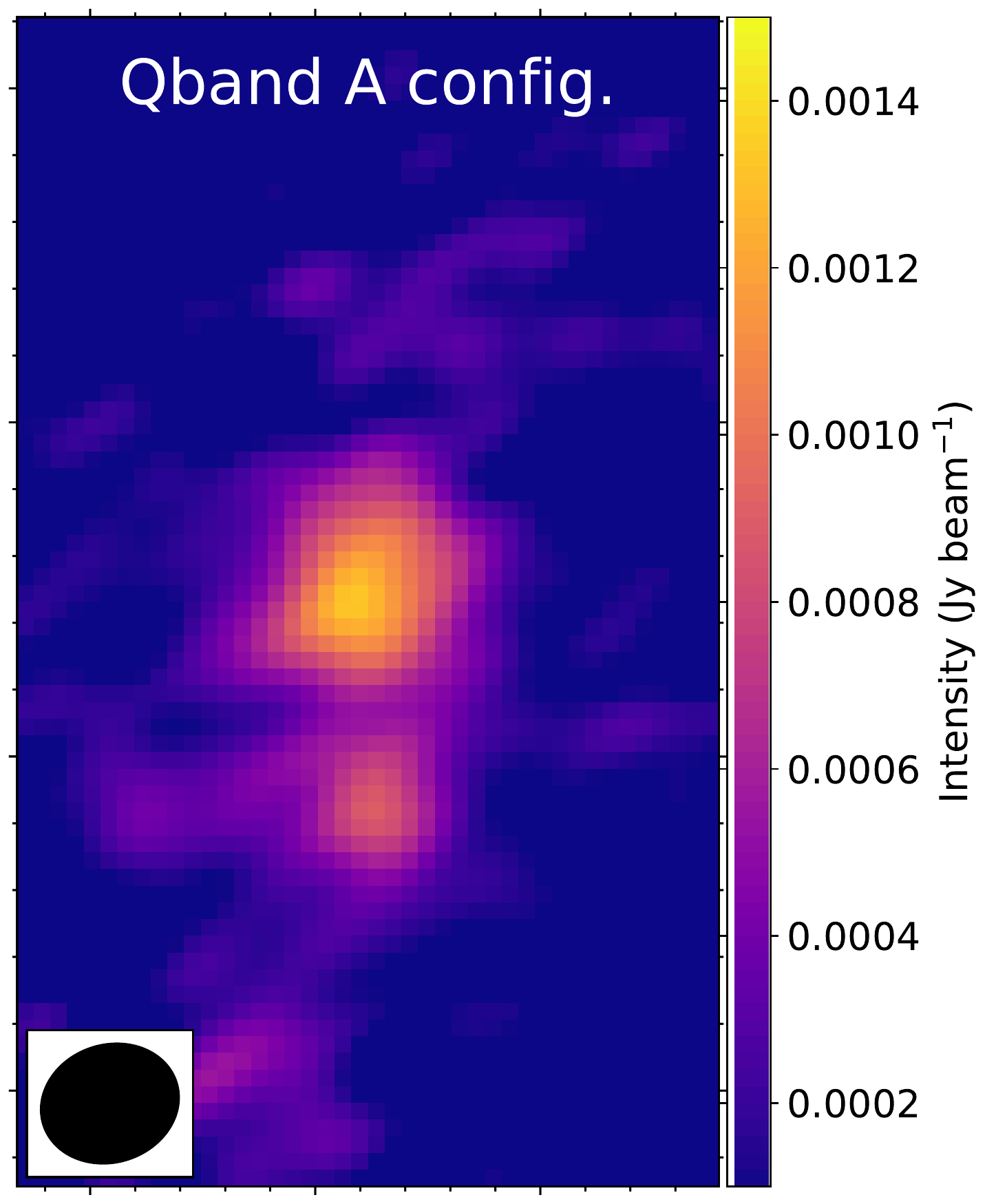}   & 
        \includegraphics[width=5.7cm]{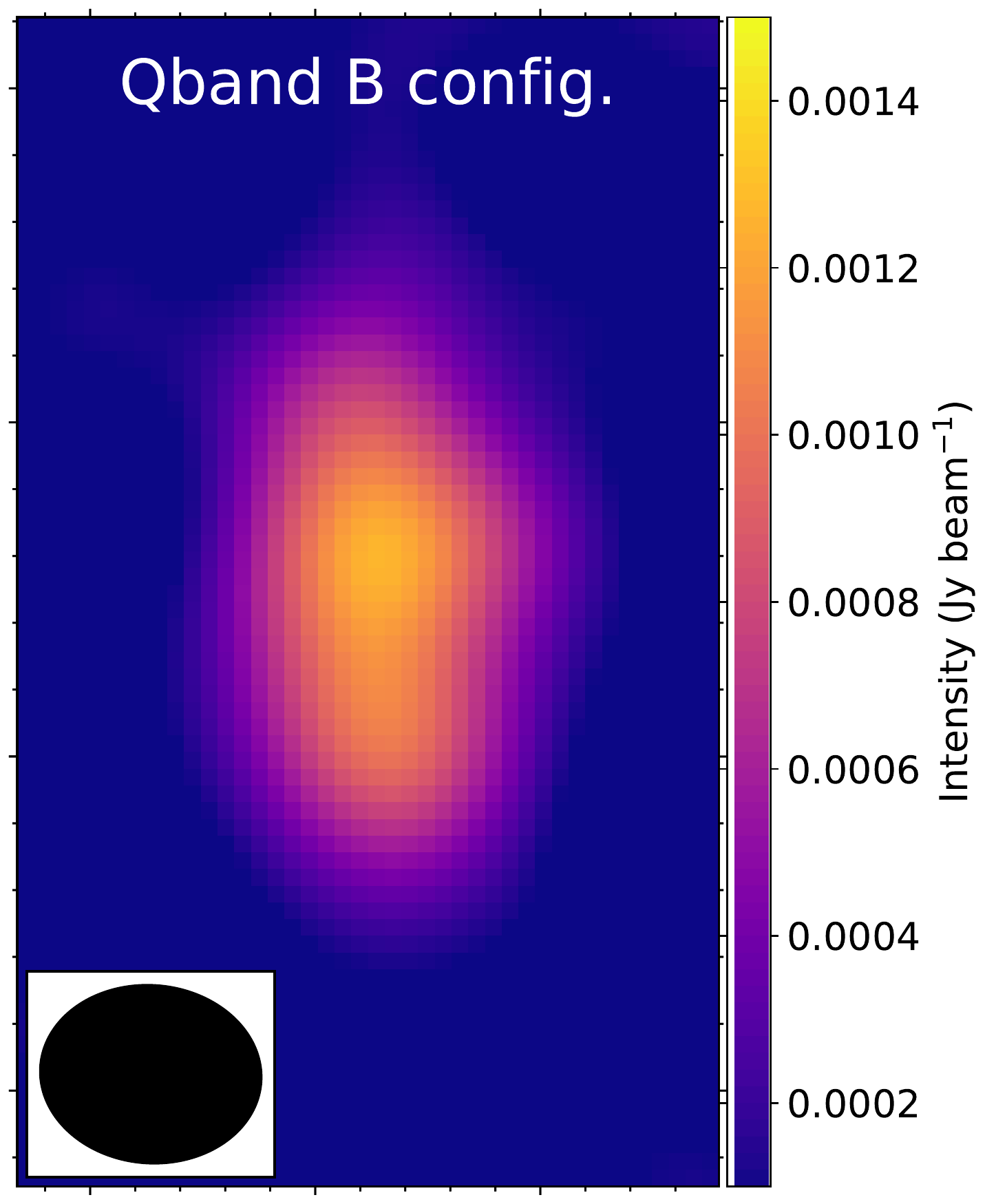}  \\
        \hspace{-1.72cm}\includegraphics[width=7.43cm]{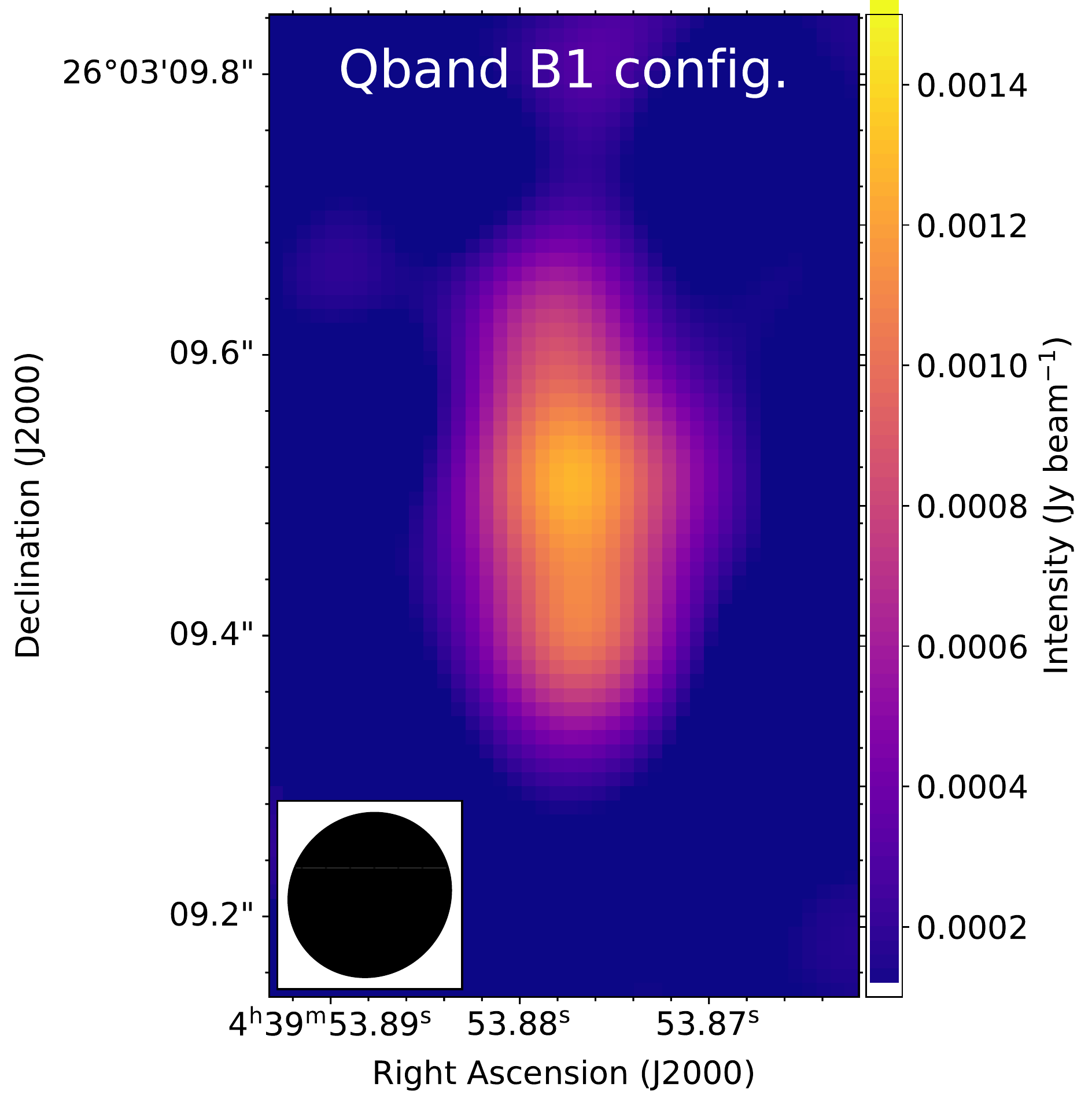}  & 
        \vspace{-7.5cm} \includegraphics[width=5.7cm]{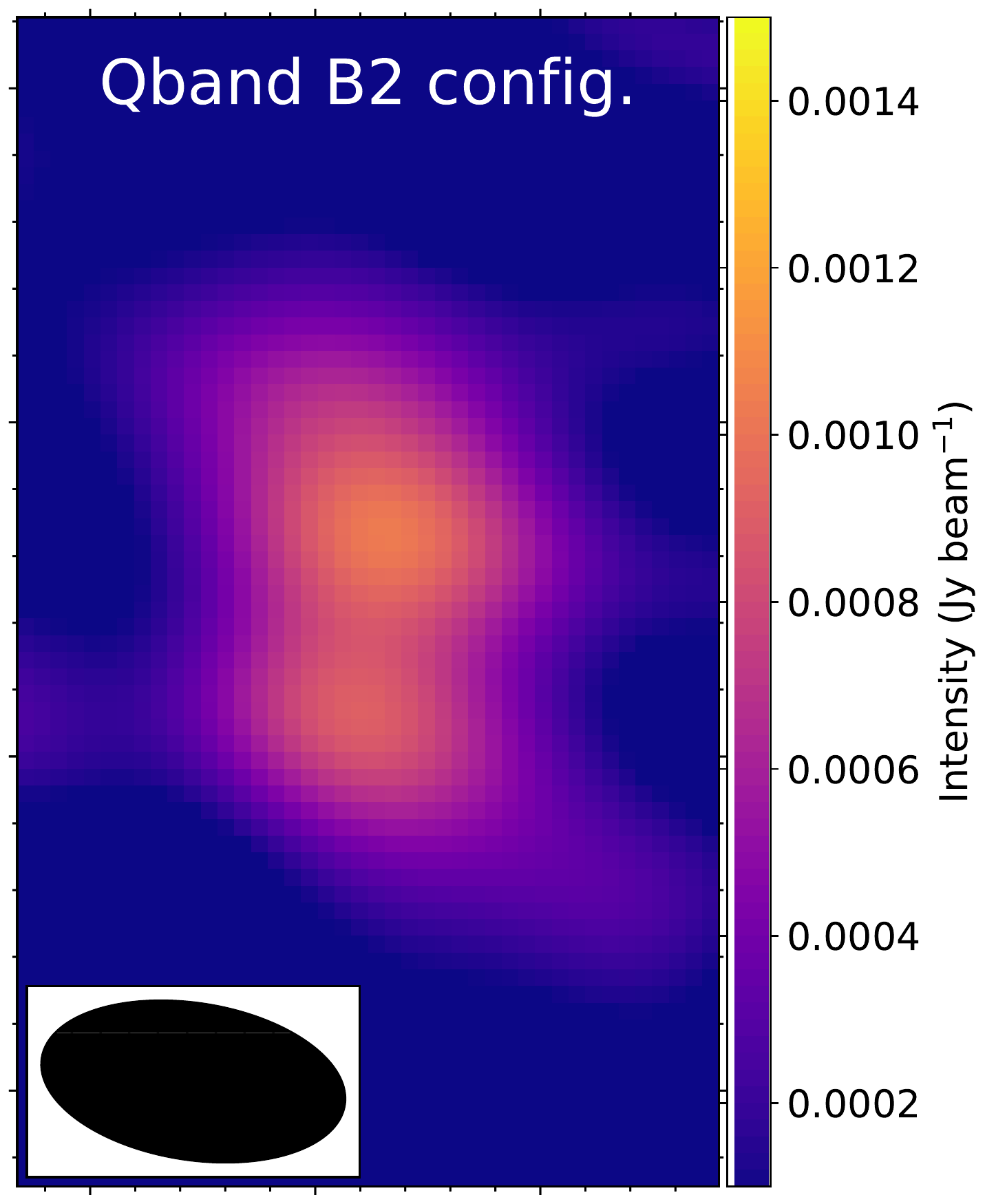}  & 
        \vspace{-7.5cm} \includegraphics[width=5.7cm]{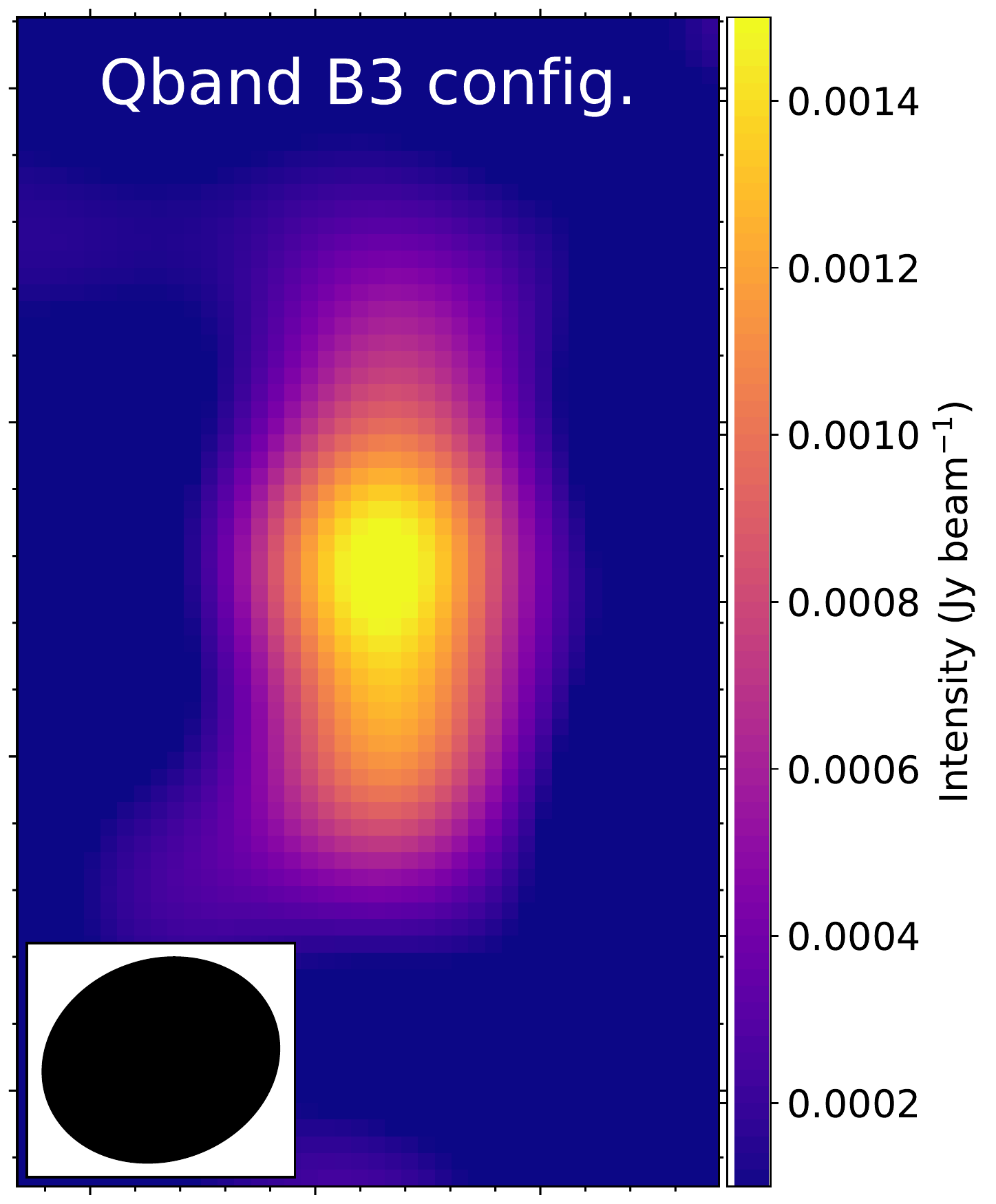}  \\
    \end{tabular}
    \caption{JVLA 22-24 GHz (K band) and 44-46 GHz (Q band) images generated with various combination of array configuration(s). These images are primary beam corrected. The details of these image are simmarized in Table \ref{tab:qbandimage}. B1, B2, and B3 refers to Tracks 4, 5, and 6, respectively.} 
    \label{fig:qbandimage}
\end{figure*}
\if0
\begin{figure*}
    \hspace{-1.cm}
    \begin{tabular}{ c c c}
        \includegraphics[width=4cm]{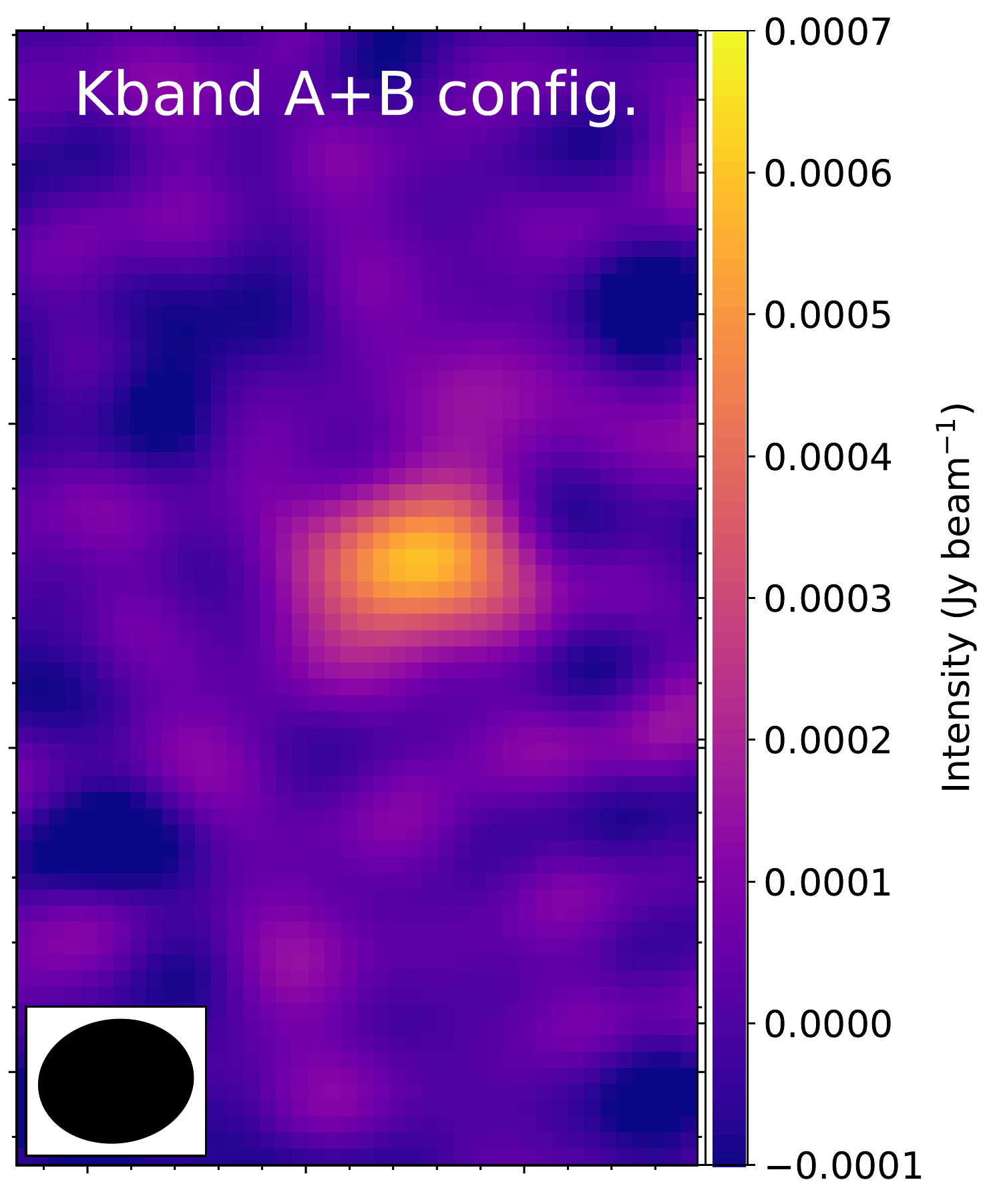} &
        \includegraphics[width=4cm]{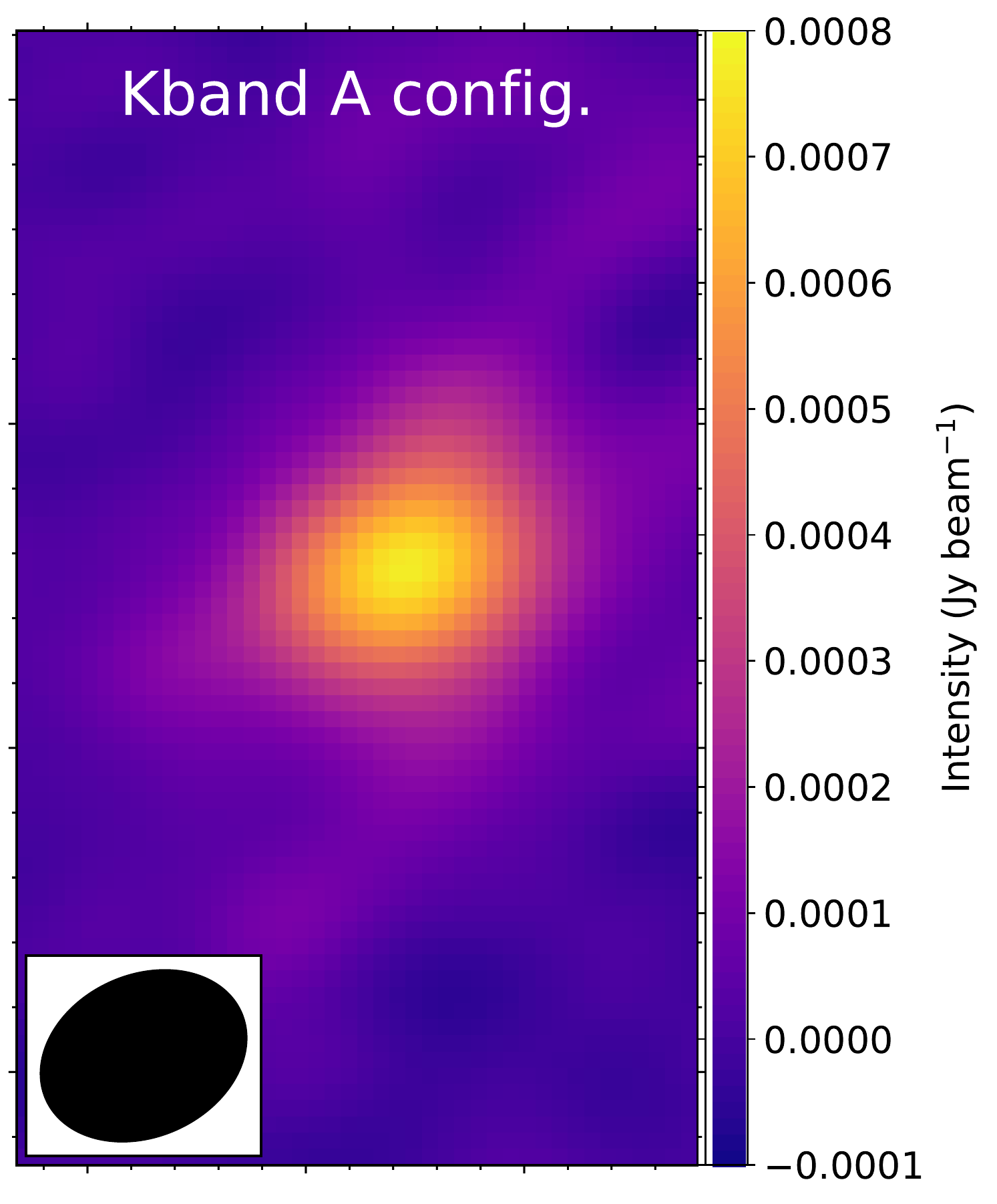} &
        \includegraphics[width=4cm]{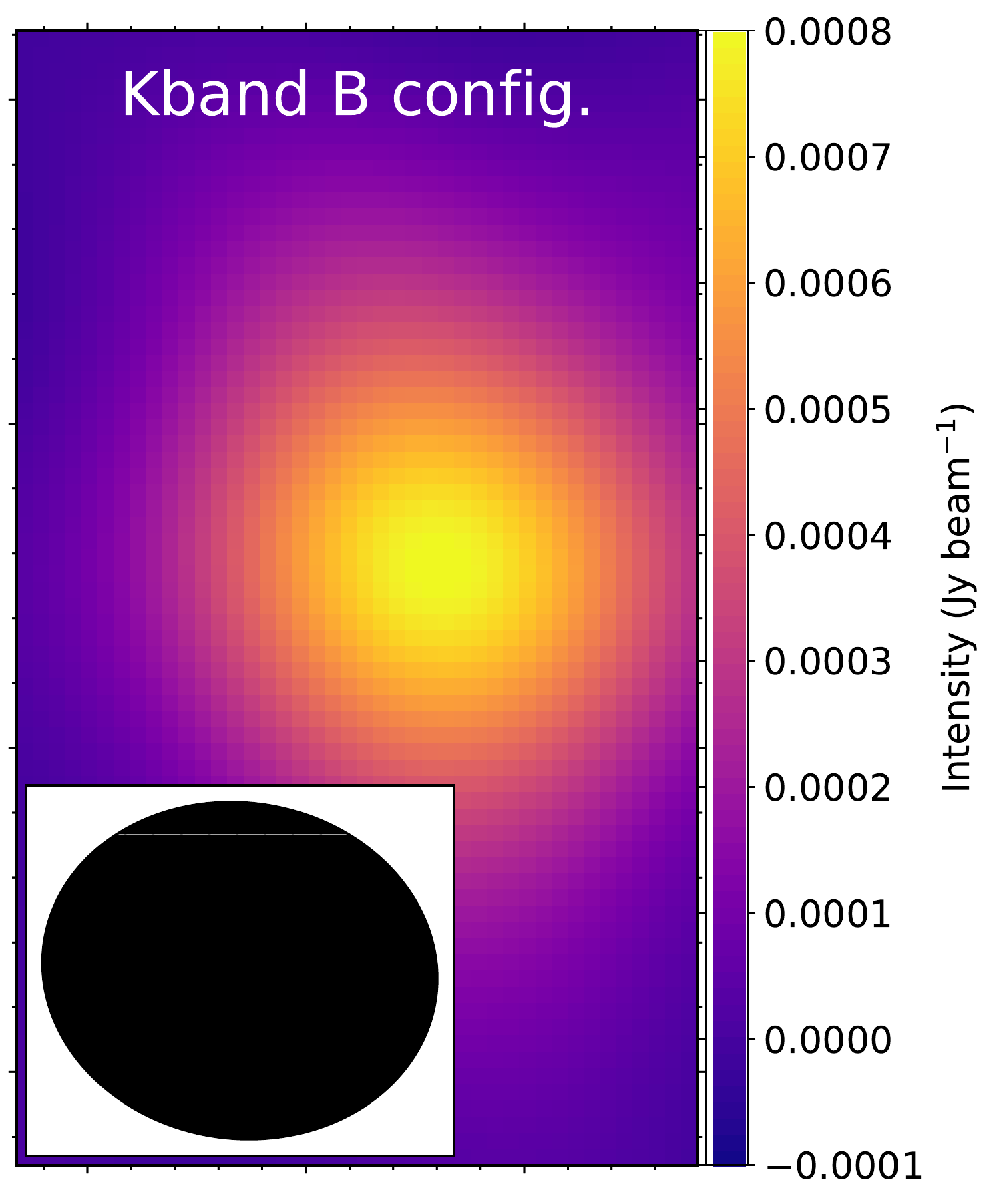} \\
        \includegraphics[width=6cm]{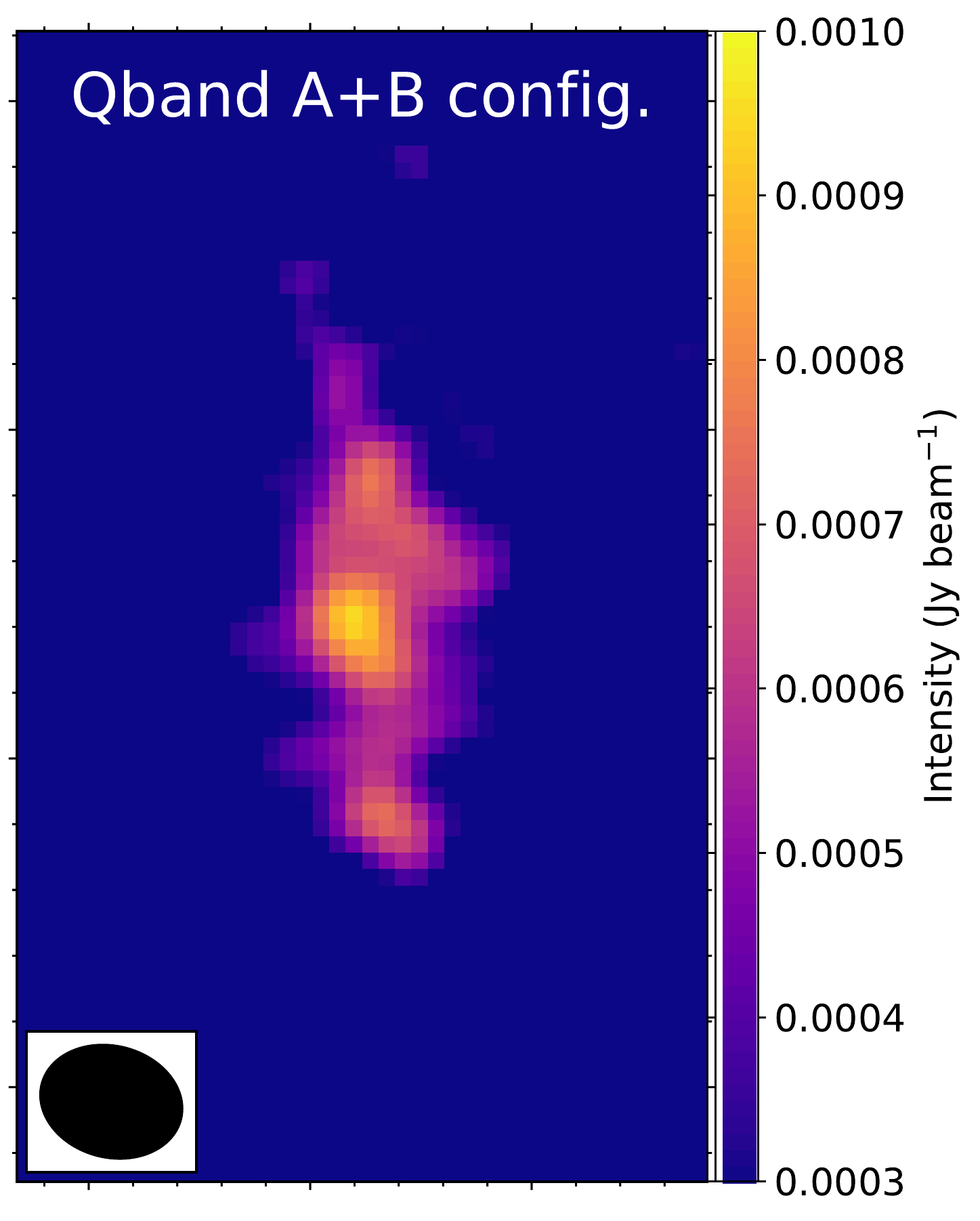} &  
        \includegraphics[width=6cm]{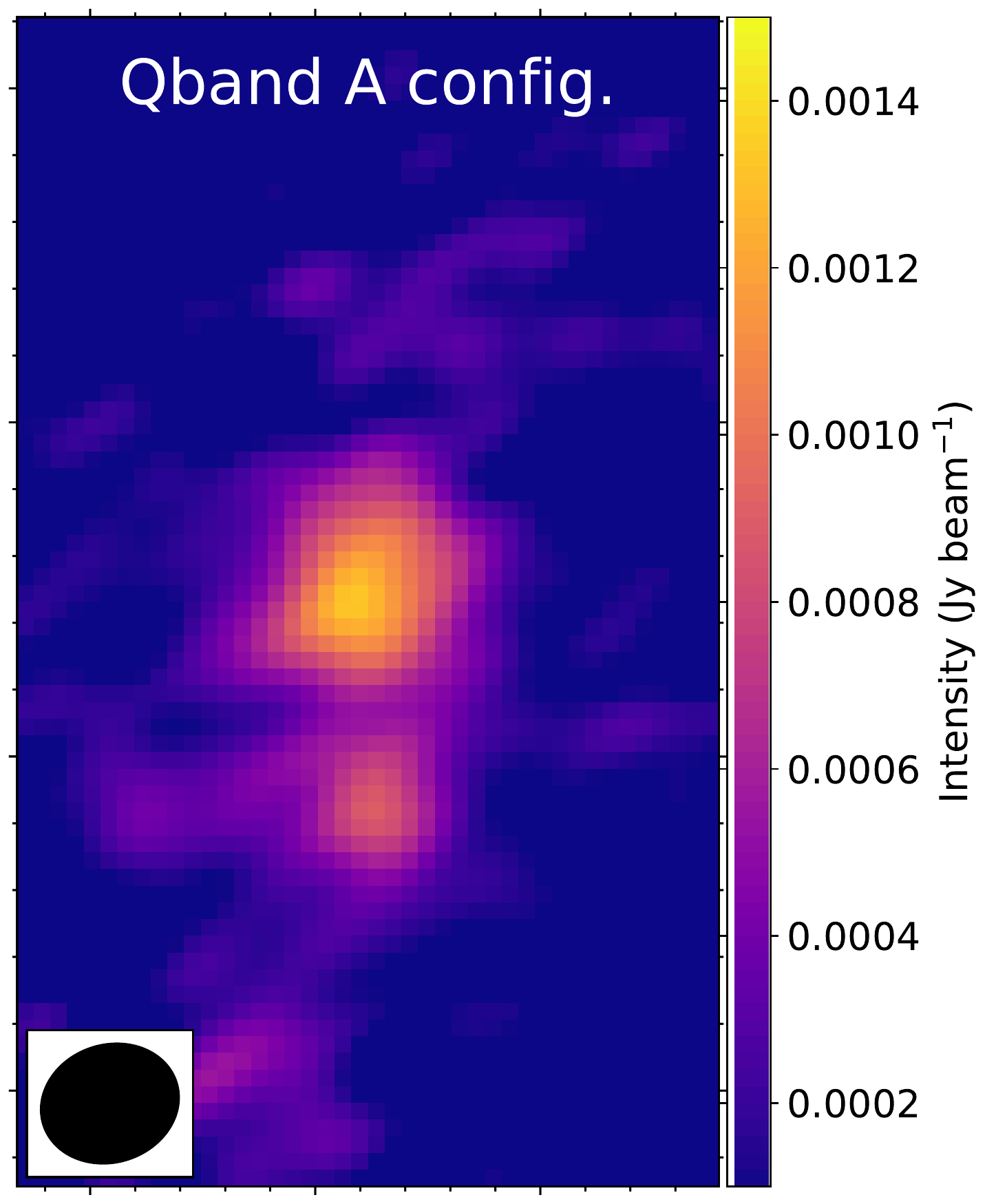}   & 
        \includegraphics[width=6cm]{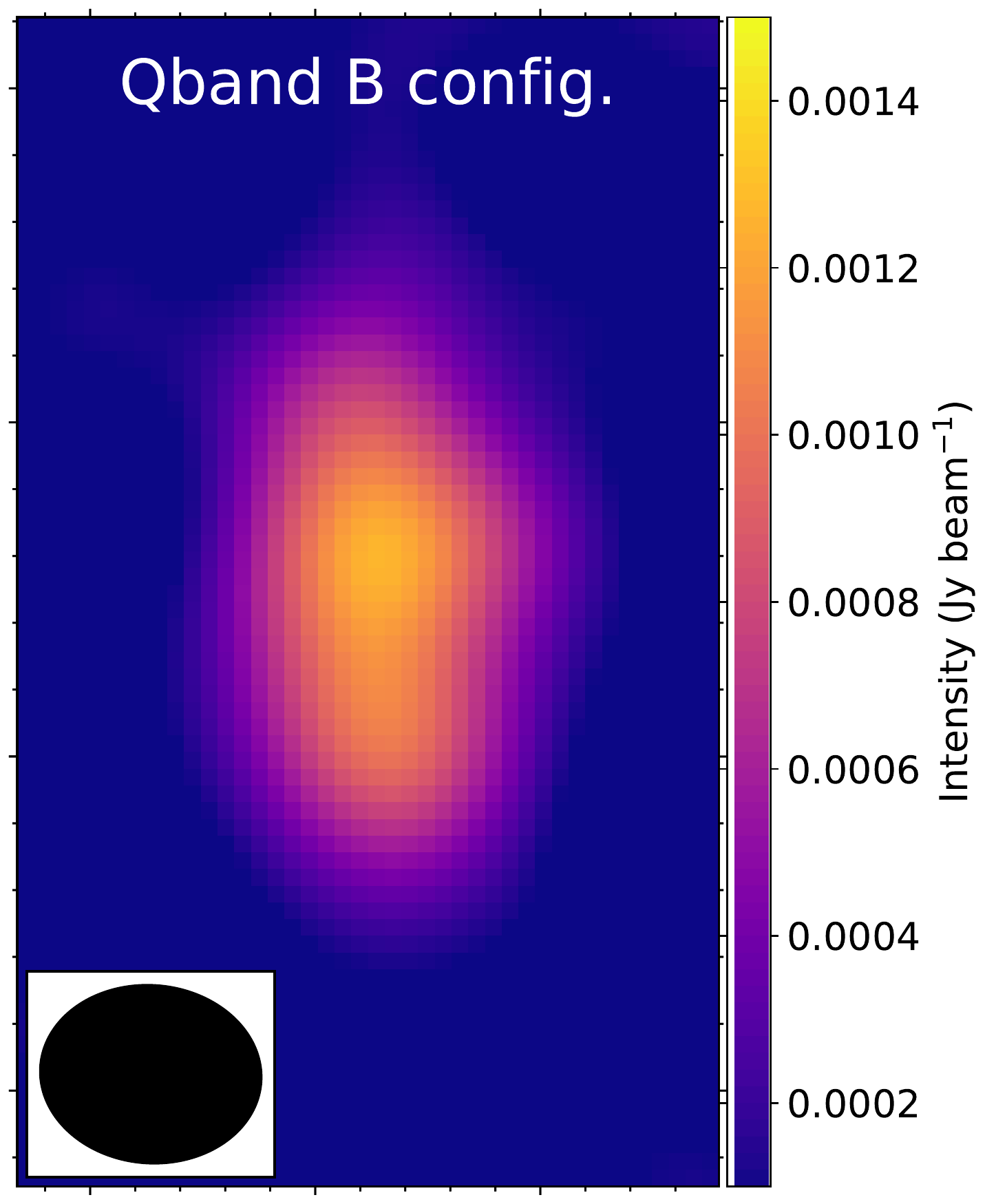}  \\
        \includegraphics[width=6cm]{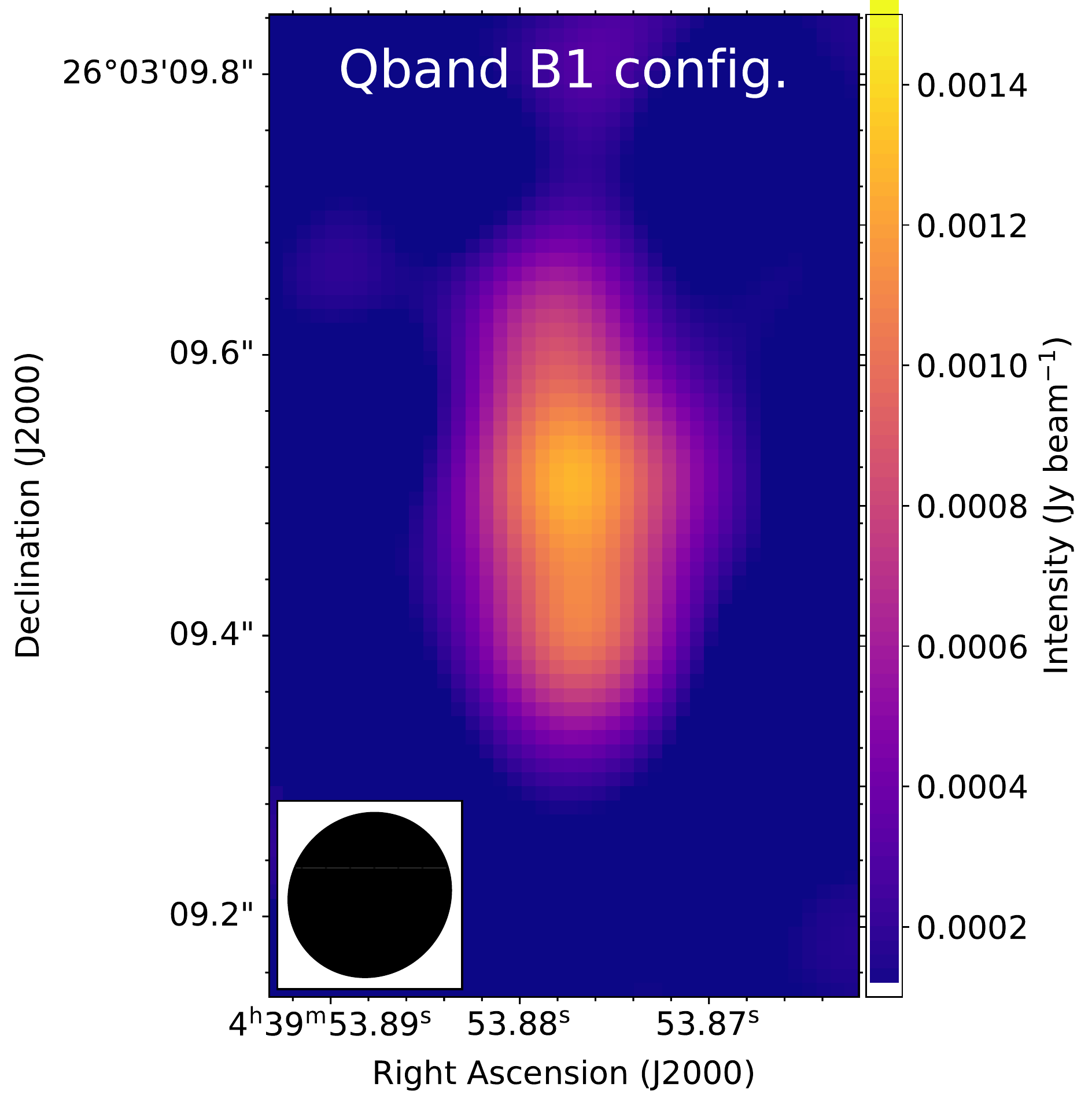}  & 
        \includegraphics[width=6cm]{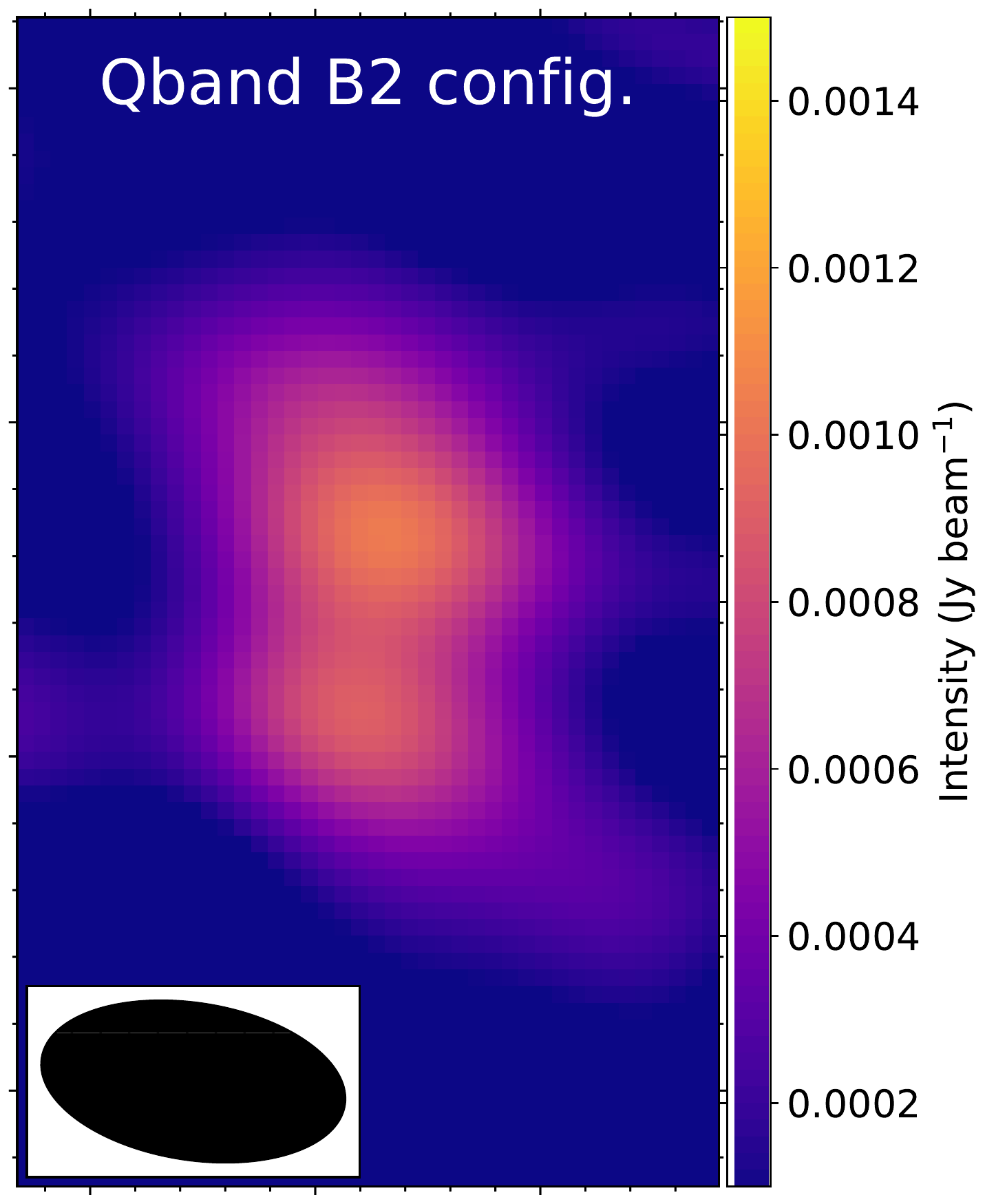}  & 
        \includegraphics[width=6cm]{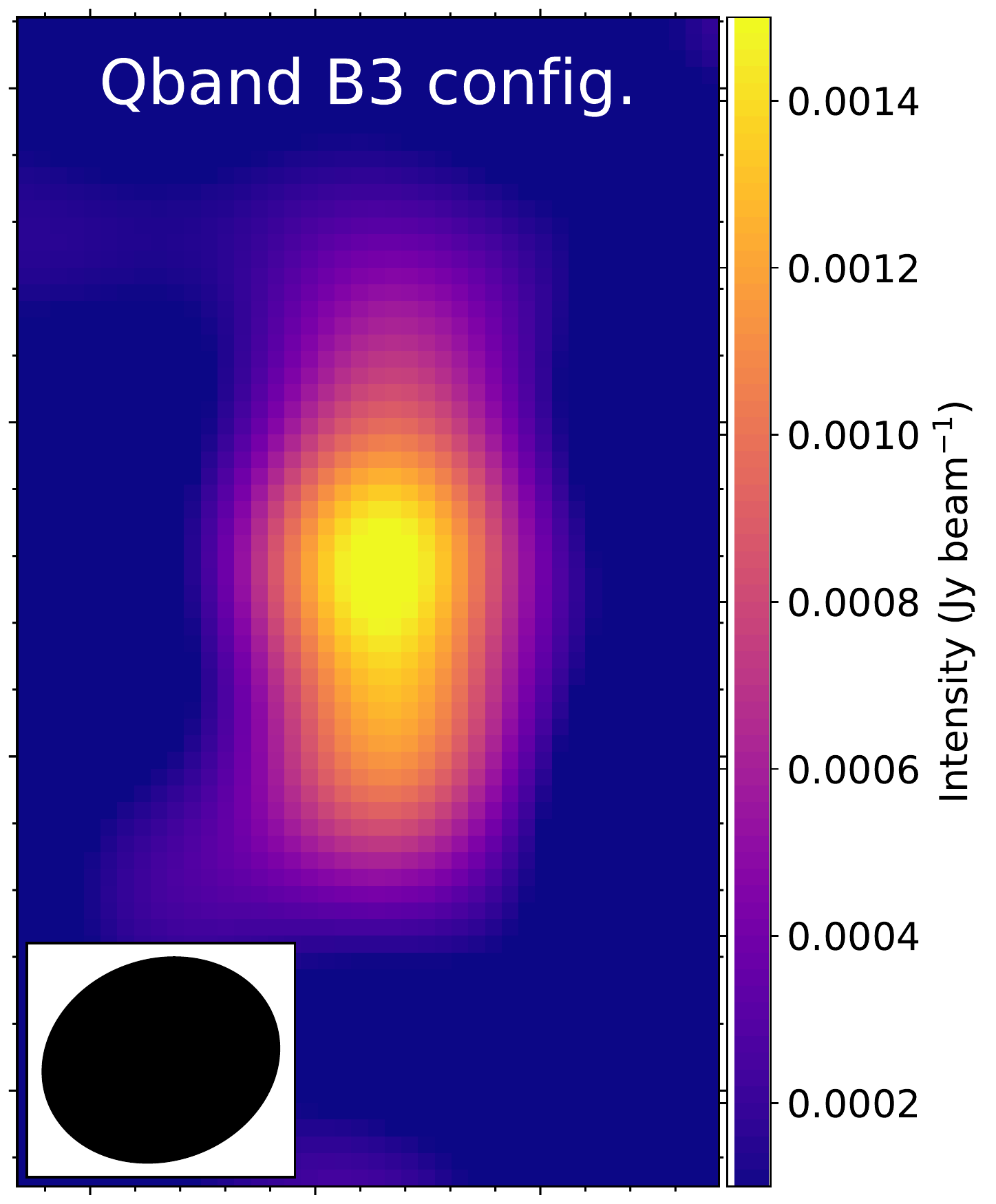}  \\
    \end{tabular}
    \caption{JVLA 44-46 GHz (Q band) images generated with various combination of array configuration(s). The details of these image are summarized in Table \ref{tab:qbandimage}. B1, B2, and B3 refers to Tracks 4, 5, and 6, respectively. 
    %\mycomment{The resolution of A+B image looks better than that of A image and the beam size shown in the panel. Why is that?}
    } 
    \label{fig:qbandimage}
\end{figure*}
\fi

We performed the multi-frequency synthesis (MFS) imaging using the CASA task \textsc{tclean}, by setting the parameter {\it nterm}$=$1.
Figure \ref{fig:qbandimage} shows the images generated from all data, all A array configuratoin (only) data, all B array configuration (only) data, and from individual of the three tracks of B array configuration observations.
The weighing scheme and the yielded synthesized beam sizes and RMS noise levels of these images are summarized in Table \ref{tab:qbandimage}.
In spite of the different synthesized beam shapes, these images consistently present an elongated (north-south) geometry, which appears clumpy and lopsided.
We recovered a $\sim$3.7 mJy total flux density at the mean frequency of $\sim$45 GHz, which is very well consistent with the 3.5 mJy flux density (at 43 GHz) reported by \citet{Li2017}, which was derived from the historical VLA observations taken in 1996-2004 \citep[c.f.][]{Loinard2002,Melis2011}.

%44-46 680-26980 J0438+3004 0.350.002
%22-24 740-28150 J0431+2037 0.680.00065
%44-46 740-31320 J0438+3004 0.360.001
%22-24 750-30100 J0431+2037 0.640.00065
%44-46 960-35200 J0438+3004 0.400.001
%22-24 930-33950 J0431+2037 0.700.0013
%40-48 210-10510 J0440+2728 0.170.0003
%18-26 200-9380 J0403+2600 1.20.00065
%40-48 170-9930 J0440+2728 0.160.0005
%18-26 200-9380 J0403+2600 1.10.00052
%40-48 160-9280 J0440+2728 0.150.0002
%18-26 150-9250 J0403+2600 0.990.00085
%\startlongtable
%\newcommand{\celln}[1]{ { \begin{tabular}{c} #1 \end{tabular}}}
\newcommand{\mulr}[1]{\multirow{2}{*}{#1}}
{  	  \movetabledown=2.4in 
\begin{rotatetable}
\begin{deluxetable*}{ c c c c c c  c c c c c c}
%\tablewidth{20cm}
  \tablecaption{Summary of observational data\label{tab:obssum}}
  \tablehead{
  \colhead{Track ID} &  \colhead{Band}& \colhead{Observing date}  & \colhead{Array config.} & \colhead{Freq. coverage}  &   \colhead{Projected baseline lengths}
  &    \colhead{Flux/Passband calib.}  &   \colhead{Phase calib.}   & \colhead{Gain calib.flux}
  & \colhead{\footnotesize Obs. ID} \\  
  (\#)    & &{\footnotesize UTC (YYYYMMDD)} &  &    {\footnotesize (GHz)}   & {\footnotesize (meters)}  &   &   & {\footnotesize (Jy)} & 
  } 
  \startdata
    \multirow{3}{*}{1}   & JVLA-C band  &   \multirow{3}{*}{20110721} &   \multirow{3}{*}{A}  &   4.1-7.8 	&  320-31800    &   \multirow{3}{*}{3C147} & J0431+2037 &   2.7$\pm$0.002 &  \\
	   & JVLA-K band 	&  &     &   22--24 	&  740--28150    &   %\mulr{3C147} 
	& J0431+2037 &   0.68$\pm$0.00065
	&evla/pdb/4128974\\	
	  	& JVLA-Q band&&&44--46&680-26980&&J0438+3004& 0.35$\pm$0.002 &(PI: Melis)\\ \hline
    \multirow{3}{*}{2}   & JVLA-C band  &   \multirow{3}{*}{20110724} &   \multirow{3}{*}{A}  &   4.1-7.8 	&  350-31840    &   \multirow{3}{*}{3C147} & J0431+2037 &   2.7$\pm$0.002 &  \\
	   & JVLA-K band  	&  &    &   22--24 	&  750-30100    &   &J0431+2037&   0.64$\pm$0.00065
		& evla/pdb/4128974\\	
	  	& JVLA-Q band&&&44--46&740-31320&&J0438+3004&0.36$\pm$0.001 &(PI: Melis)\\ \hline	
    \multirow{3}{*}{3}   & JVLA-C band  &   \multirow{3}{*}{20110806} &   \multirow{3}{*}{A}  &   4.1-7.8 	&  660-35690    &   \multirow{3}{*}{3C147} & J0431+2037 &   2.7$\pm$0.003 &  \\
	   & JVLA-K band 	&   &     &   22--24 	&  930-33950    &   &J0431+2037 &  0.70$\pm$0.0013 
	& evla/pdb/4128974\\	
	  	& JVLA-Q band&&&44--46&960-35200&&J0438+3004& 0.40$\pm$0.001 &(PI: Melis)\\ \hline
    \multirow{3}{*}{4}   & JVLA-C band  &   \multirow{3}{*}{20131004} &   \multirow{3}{*}{B}  &   4.1-7.8 	&  220-10910    &   \multirow{3}{*}{3C147} & J0403+2600 &   2.6$\pm$0.003 &  \\
	   & JVLA-K band 	&   &   &   18--26 	&  200-9380    &    &J0403+2600 &   1.2$\pm$0.00065
	& evla/pdb/21340702\\	
	  	& JVLA-Q band&&&40--48&210-10510&&J0440+2728&0.17$\pm$0.0003 &(PI: Melis)\\ \hline
    \multirow{3}{*}{5}   & JVLA-C band &   \multirow{3}{*}{20131019} &   \multirow{3}{*}{B}  &   4.1-7.8 	&  160-10080    &   \multirow{3}{*}{3C147} & J0403+2600 &   2.6$\pm$0.004 &  \\
	   & JVLA-K band 	&   &   &   19--26  &       200-9380 &   &J0403+2600  &   1.1$\pm$0.00052
	&evla/pdb/21340702\\	
	  	& JVLA-Q band&&&40--48& 170-9930 &&J0440+2728&0.16$\pm$0.0005 &(PI: Melis)\\ \hline	
    \multirow{3}{*}{6}   & JVLA-C band  &   \multirow{3}{*}{20131110} &   \multirow{3}{*}{B}  &   4.1-7.8 	&  140-10240    &   \multirow{3}{*}{3C147} & J0403+2600 &   2.6$\pm$0.003 &  \\
	   & JVLA-K band 	&   &   &   18--26 	&  150-9250     &    &J0403+2600 &   0.99$\pm$0.00085
	& evla/pdb/21340702\\	
	  	& JVLA-Q band&&&40--48&160-9280&&J0440+2728& 0.15$\pm$0.0002 &(PI: Melis)\\ \hline	
	\mulr{7}   & \mulr{ALMA-Band 3}	&  \mulr{20171113}   	& 	\mulr{C43-8}     	&  \mulr{85--100} & \mulr{113-13900} &   \mulr{J0510+1800}  &\mulr{J0435+2532}	& \mulr{0.090$\pm$0.0006}
	&A001/X1220/X5d2 \\    
		  	&&&&&&&& &(PI: Sakai)\\ \hline	
	\mulr{8}   & \mulr{ALMA-Band 3}	&  \mulr{20171114}	&    	\mulr{C43-8}	&  \mulr{85--100} 	&  \mulr{113-12300}     &   \mulr{J0510+1800}  &\mulr{J0435+2532} 	& \mulr{0.091$\pm$0.0005 }  
	&A001/X1220/X5d2 \\   
		  	&&&&&&&& &(PI: Sakai)\\ \hline		
	\mulr{9}   & \mulr{ALMA-Band 3}	&  \mulr{20171114} 	&  	\mulr{C43-8}    	&  \mulr{85--100} 	& \mulr{113-12300}      &   \mulr{J0510+1800}  &	\mulr{J0435+2532}	& \mulr{0.091$\pm$0.0005}  
	&A001/X1220/X5d2\\   
			  	&&&&&&&& &(PI: Sakai)\\ \hline	
	\mulr{10}   & \mulr{ALMA-Band 4}	&  \mulr{20161119}    	& 	\mulr{C40-4}  	&	\mulr{138--150} 	&\mulr{15-704}           & \mulr{J0510+1800}  &  \mulr{J0438+3004}  &   \mulr{0.35$\pm$0.003 }
	& A001/X5ac/Xe4a \\	
			  	&&&&&&&& &(PI: Oya)\\ \hline		
	\mulr{11}   & \mulr{ALMA-Band 4}	&  \mulr{20170903}   	&  	\mulr{C40-7} 	&	\mulr{138--151} 	&\mulr{21-3700}          & \mulr{J0510+1800}  &  \mulr{J0440+2728}  &   \mulr{0.11$\pm$0.006   }
	& A001/X5ac/Xe4a\\	
			  	&&&&&&&& &(PI: Oya)\\ \hline			
	\mulr{12} & \mulr{ALMA-Band 7}	&20150718	& \mulr{C34-7(6)}	& \mulr{338--352}  &   \mulr{42-1574}    	&J0423-013/J0423-0120  	&\mulr{J0438+3004} 	&    	0.12$\pm$0.004
	&A001/X10f/X868\\ 
			  	&&20150720&&&&J0423-013/J0433+0521&& 0.12$\pm$0.001 &(PI: Sakai)\\ \hline				
	\mulr{13}  & \mulr{ALMA-Band 7}	&20170729	& C40-5 & \mulr{338--352}   &  17--1100&\mulr{J0510+1800}	&J0438+2728	 & 0.42$\pm$0.04
	&A001/X8aa/X23\\
			  	&&20170905&C40-8&&168--6800&&J0438+3004& 0.20$\pm$0.005 &(PI: Sakai)%\\ \hline				
  \enddata
  \tablecomments{
  %All the listed tracks observed quasar 3C147 for absolute flux and passband calibrations. 
  The gain calibrators flux for ALMA Band~3, 4, and 7 are taken at 
  $87 \GHz$ ($1.875 \GHz$ bandwidth),
  $150.4\GHz$ ($58.59 {\rm \, MHz}$ bandwidth), 
  and $345.8\GHz$, respectively. 
  The errors of the gain calibrator flux show statistical uncertainties but absolute flux uncertainties. 
 }
\end{deluxetable*}
\end{rotatetable}
}

\begin{deluxetable*}{ c c c r c c c }

%  \tablecaption{Summary for Q band (44-46 GHz; 6.5-6.8 mm) images\label{tab:qbandimage}}
\tablecaption{Summary for JVLA images\label{tab:qbandimage}}
  \tablehead{
  \colhead{Label}    & Included observing tracks &  Briggs robust parameter &  \colhead{Synthesized beam}  & RMS noise level & Recovered flux & \colhead{Peak intensity} \\
                   &                                        &                                   &  (\beam; P.A.)
                   & (mJy\,beam$^{-1}$)            & (mJy) & (mJy\,beam$^{-1}$)
  }
  \startdata
    C band &  \#1-6 & -2 &  0\farcs26$\times$0\farcs23; -65$^{\circ}$   &  0.027 & 0.65 & 0.71 \\ % center x: 04:39:53.879  center y: 26.03.09.478, 4.41036 times 4.32193 with ellipse for C and K
      K band  	& \#1--6   		& -2  	& 0\farcs095$\times$0\farcs075; -82$^{\circ}$    &  0.066    &  1.1 & 0.59\\
%KBand 0.095028154552 0.0753232911229 -82.9285049439 6.66726964383e-05 0.0007771388122534972 0.0005922382
      Q band %A+B config. 	
      & \#2--6  		& -2  	& 0\farcs087$\times$0\farcs068; 76$^{\circ}$    &  0.11    & 3.7   &	0.96\\
%QBand 0.0873247683048 0.0678472742438 75.8460464477 0.000107937851917 0.0036544828337241597 0.0009600492      
      Band 3 			& \#7--9   		& 0.5  	& 0\farcs155$\times$0\farcs068; -2$^{\circ}$    &  0.020    &   23 &5.4\\
%Band3 0.15459702909 0.0679504796862 -1.59628295898 0.000170770803508 0.020769210448065678 0.0054653906      
      Band 4			& \#10--11   	& -2  	& 0\farcs195$\times$0\farcs133; -5.2$^{\circ}$    &  0.15    &   73&  1.6\\
%Band4 0.195180743933 0.132556825876 -5.15666198731 0.000739700087702 0.05635684936243024 0.016383363           
      Band 7 			& \#12--13   	& 0.5  	& 0\farcs072$\times$0\farcs067; -11$^{\circ}$    &  0.15    &  370 & 1.9 \\ \hline \hline
%Band7 0.0715738758445 0.0668535903096 -10.8800077438 0.000848007745751 0.34149295533870827 0.018992592
    C band A+B config. & \#1--6 & -2    & 0\farcs26$\times$0\farcs23; -65$^{\circ}$ & 0.027    &   0.65 & 0.71 \\
    C band A config. & \#1--3 & 2       & 0\farcs26$\times$0\farcs23; -65$^{\circ}$ & 0.028 & 0.81 & 0.72\\
    C band B config. & \#4--6 & -2      & 0\farcs72$\times$0\farcs64; 78$^{\circ}$ & 0.031 & 0.50 &  0.51 \\\hline
        K band A+B config. & \#1--6 & -2    & 0\farcs095$\times$0\farcs075; -83$^{\circ}$ & 0.066    &   1.1 & 0.59\\
    K band A config. & \#1--3 & 2       & 0\farcs13$\times$0\farcs099; -66$^{\circ}$ & 0.036 & 2.1 & 0.77 \\
    K band B config. & \#4--6 & -2      & 0\farcs24$\times$0\farcs21; 82$^{\circ}$ & 0.020 & 1.0 & 0.81 \\\hline
    Q band A+B config. 	& \#2--6  		& -2  	& 0\farcs087$\times$0\farcs068; 76$^{\circ}$    &  0.11    & 3.7   & 0.96\\ 
    Q band A config.   & \#2, 3   &  0  & 0\farcs084$\times$0\farcs070; -72$^{\circ}$   &  0.13    & 3.7 &  1.4\\
    Q band B config.   & \#4--6   &  -2   & 0\farcs13$\times$0\farcs11; 86$^{\circ}$      &  0.061 & 3.7   &1.3 \\
    Q band B1 config.  & \#4      &  -2   & 0\farcs12$\times$0\farcs11; -40$^{\circ}$     &  0.10    & 3.6  & 1.3  \\
    Q band B2 config.  & \#5      &  -2   & 0\farcs18$\times$0\farcs094; 81$^{\circ}$     &  0.098   & 3.7  &  1.1 \\
    Q band B3 config.  & \#6      &  -2   & 0\farcs14$\times$0\farcs12; -71$^{\circ}$     &  0.10    & 3.8     &1.6\\ 
  \enddata
%  \tablecomments{These images were created using only the 44-46 GHz passband to permit robustly combining the data taken with A and B array configurations. More details of individual observing tracks are provided in Table \ref{tab:obssum}. We quote a nominal 10\% absolute flux error.
  %}
\end{deluxetable*}

\section{Spectral Index $\alpha$ \& Dust growth}    \label{sec:alphaindex}

\begin{figure*}[htbp]
    \begin{center}
	\includegraphics[clip, width = \linewidth]{%figs/
	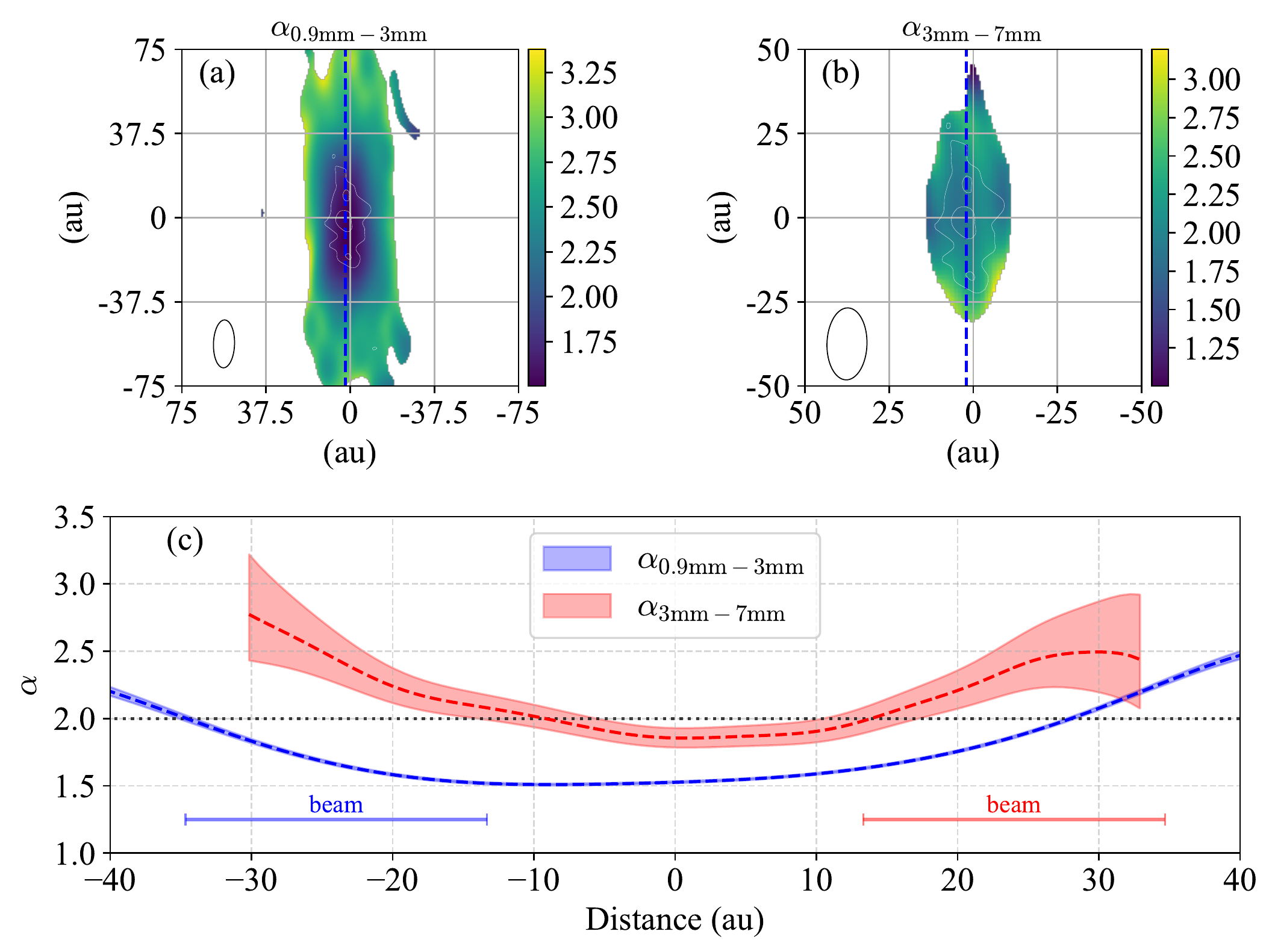}
    \caption{
    (a,b) maps for $\alpha_{\rm 0.9\mm-3\mm}$ and $\alpha_{\rm 3\mm-7\mm}$, respectively.
    The synthesized beam is shown by the ellipses at the bottom left in each panel.
    The beam sizes are 0\farcs16$\times$0\farcs069;-1.6$^{\circ}$ 
    and 0\farcs16$\times$0\farcs087;-1.6$^{\circ}$ for (a) and (b), respectively.
    The white contours are plotted in the same manner as the blue contours in the JVLA images of \fref{fig:images}.
    The blue dashed lines indicate the midplane. 
    (c) sliced distributions of $\alpha$ along the midplane. The shaded regions represent $\pm 1 \sigma$. 
    The dotted black line indicates $\alpha =  2$ for reference. 
    The sliced beam sizes ($\sim 21\au$ for both) are shown 
    at bottom left and bottom right for (a) and (b), respectively.
    }
    \label{fig:alphamaps}
    \end{center}
    \end{figure*}

	We derive $\alpha$ with the highest-resolution data:
	Band~7, Band~3, and Q~band (\fref{fig:alphamaps}). 
	The values of $\alpha_\text{3mm-7mm}$ are
	$\sim 2.49_{\,-0.51} ^{\,+0.60}$ 
	and 
	$2.77_{\,-0.54} ^{\,+0.67}$ at $d = 30\au$ to the north and to the south, respectively, 
	where $d$ is the distance from clump-C.
	The derived value is consistent with that previously reported for Band 7 and Band 6 ($1.3\mm$),
	$\alpha _\text{0.7mm-1.3mm} \sim 2.7$ \citep{Sakai2019}.
	
	The profile of $\alpha_\text{0.9mm-3mm}$ is symmetric with respect to the center. 
	The typical value is $\alpha_\text{0.9mm-3mm} = 2.5$ at $d = 40\au$ and decreases towards the center. 
	It reaches $\alpha_\text{0.9mm-3mm} = 2$ at $d = 30\au$.
	%	 ($\alpha = 2.08_{\,-0.15} ^{\,+0.15}$ at $d = 30\au$ to the north; $\alpha = 1.84_{\,-0.17} ^{\,+0.17}$ at $d = 
	The value of $\alpha_\text{0.9mm-3mm}$ is apparently small ($< 2$) for $ d < 30 \au$,
	which results from the lower brightness temperature at Band~7 than at Band~3.

	One possible explanation for the anomalously low $\alpha$ is observing different radii 
	at various frequencies \citep{Li2017, Galvan2018}. 
	Since opacities are larger for shorter wavelengths, 
	hot dust layers can be obscured by forefront dust at larger radii in our edge-on disk.
	Lower temperatures are expected for shorter-wavelength bands in this case.
	The Band 4 image shows $T_{\rm b}$ nearly equal to that at Band 7
	despite the relatively large beamsize.
	Hence, this obscured hot dust model is qualitatively consistent with our ALMA images.

	An alternative explanation for an anomalously low $\alpha$ is dust scattering effects. 
	Our source is likely optically thick, 
	and the maximum grain size, $a_{\rm max}$, appears to be $ a_{\rm max} \lesssim 100  \, {\rm \upmu m}$ ($\alpha < 3$; \fref{fig:alphamaps}).
	The frequency variation of dust albedo yields a dimmed thermal emission at certain wavelengths under such condition \citep{Liu2019,Zhu2019}.
	Note that 
	dust polarization structure have found to be consistent with dust scattering
	in the disk of L1527~IRS \citep{Segura-Cox2015, Harris2018}.
	The above two possibilities to explain the anomalous $\alpha$ are not mutually exclusive. 
	Both of them can work in our case.

    \section{Fitting results}
    \begin{figure*}
    %r.nakatani:/Volumes/PromisePegasus/L1527/compare/image_casa> python ~/Dropbox/Liouhei_Ryohei/PythonScripts/fitsimaging_via_astropy.py --2dfit QBand/*pbcor.regrid-temp-Band4subimage.fits
        \centering
        \includegraphics[width = \linewidth/2, clip]{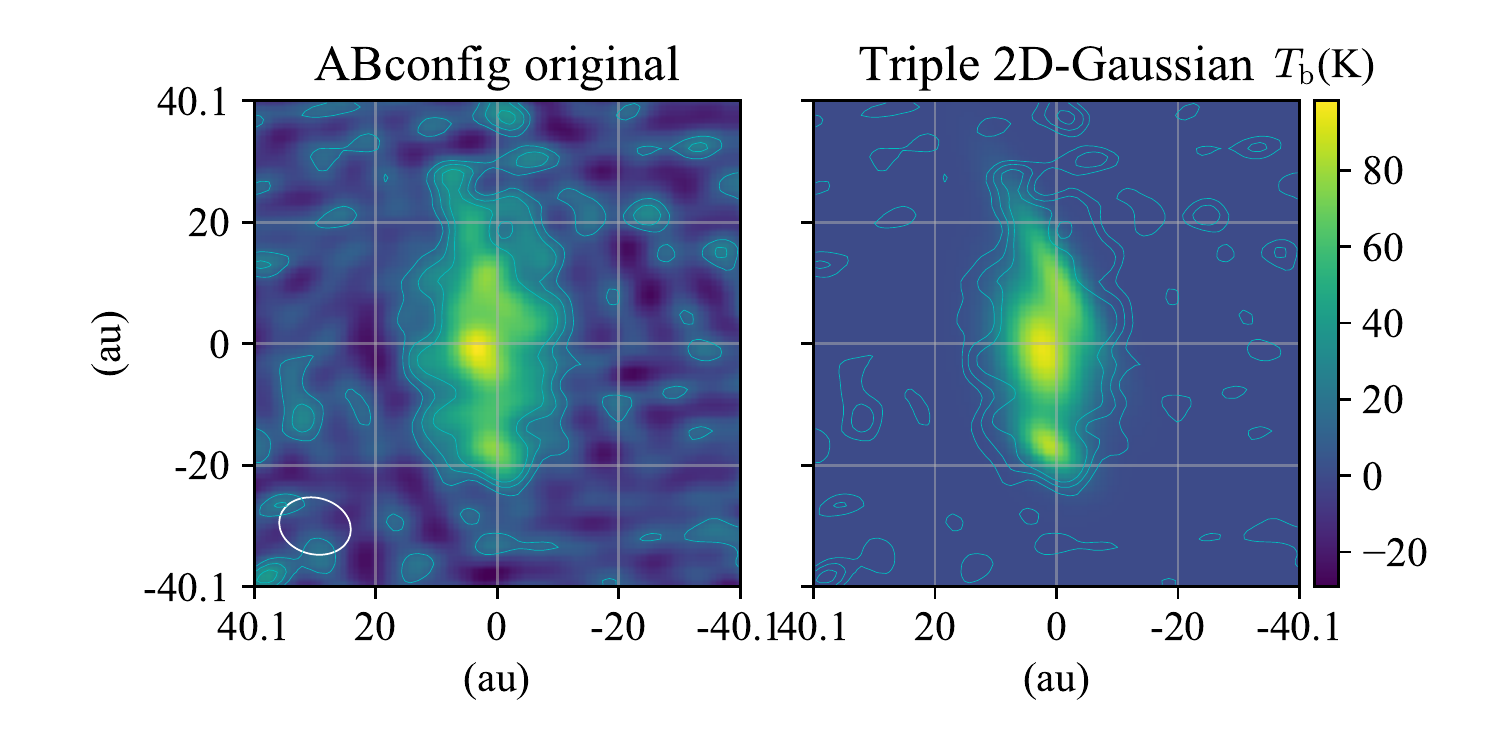}\includegraphics[width = \linewidth/2, clip]{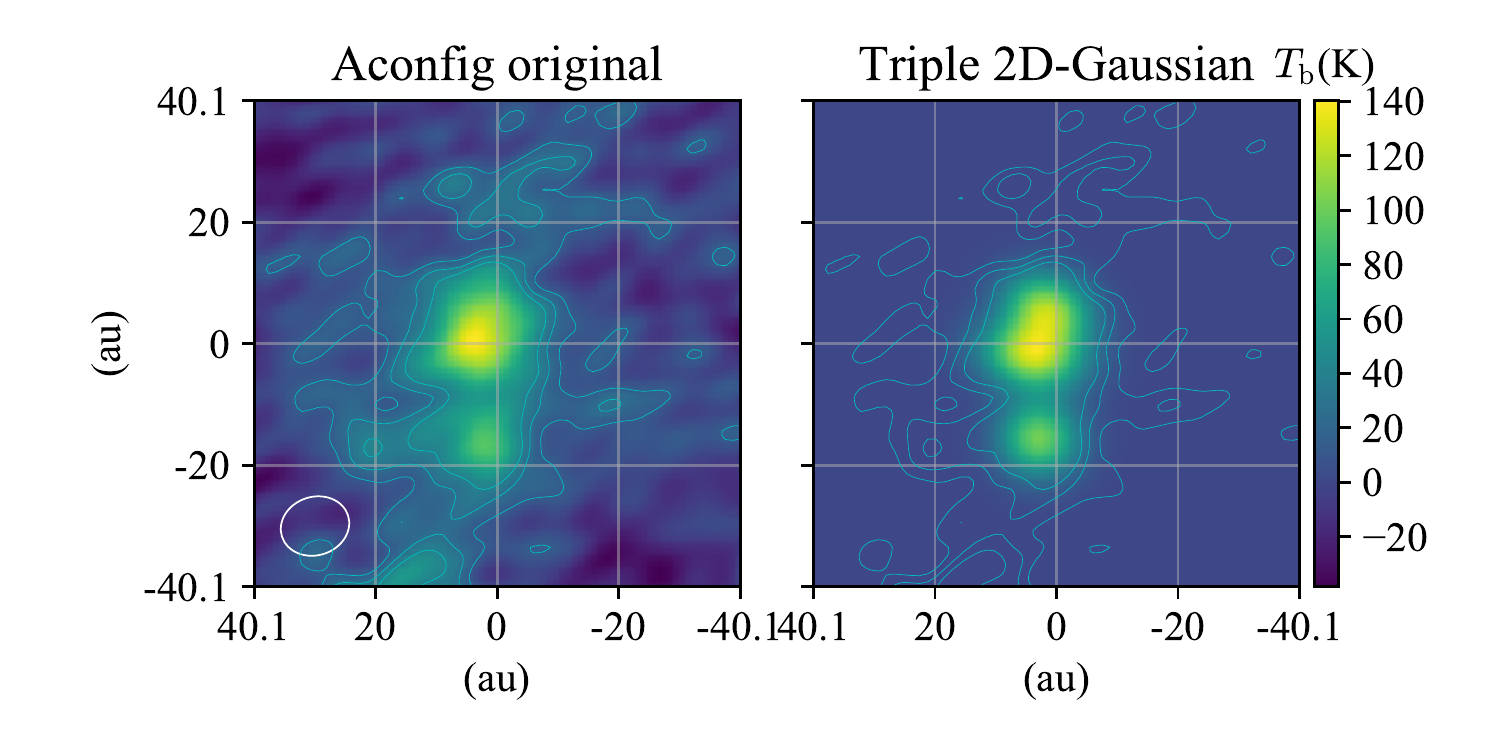}
        \includegraphics[width = \linewidth/2, clip]{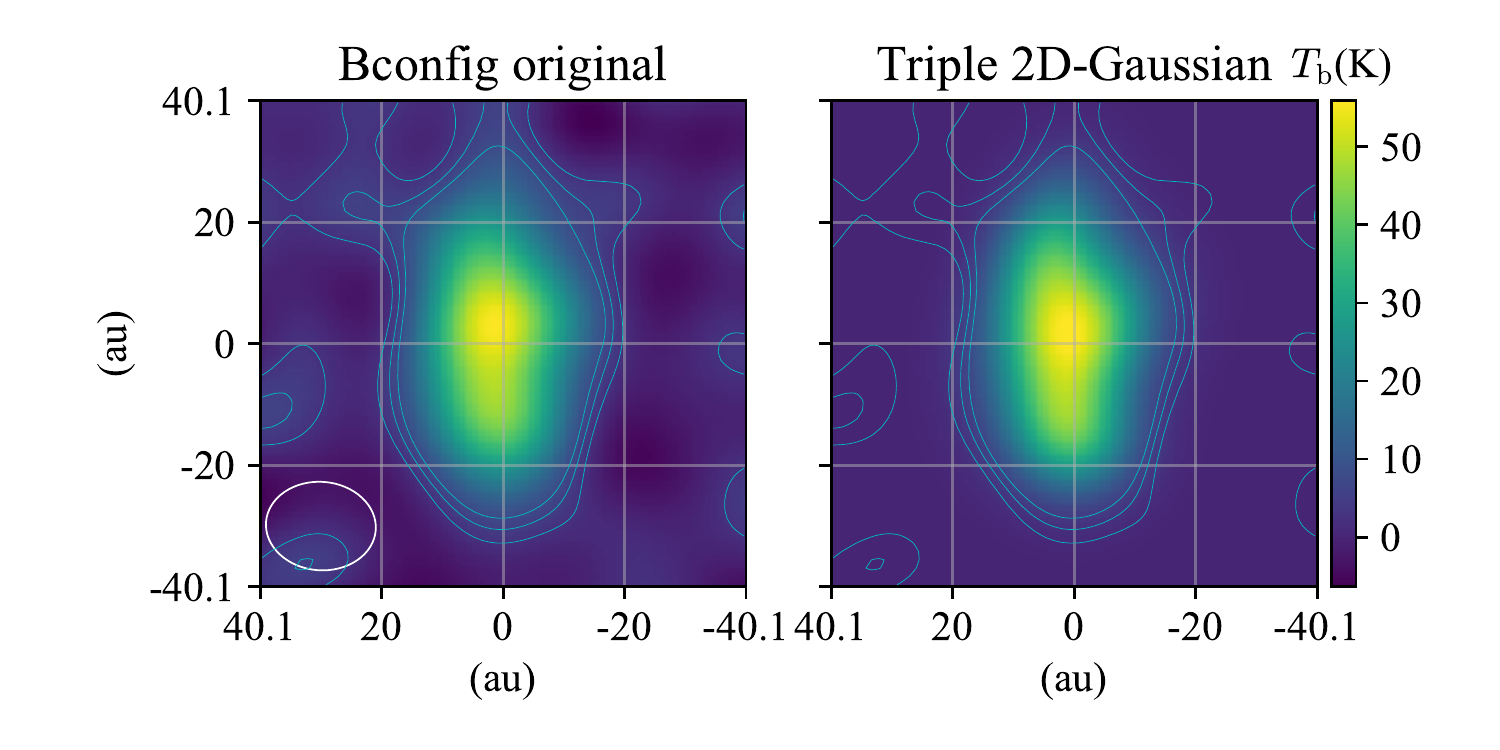}\includegraphics[width = \linewidth/2, clip]{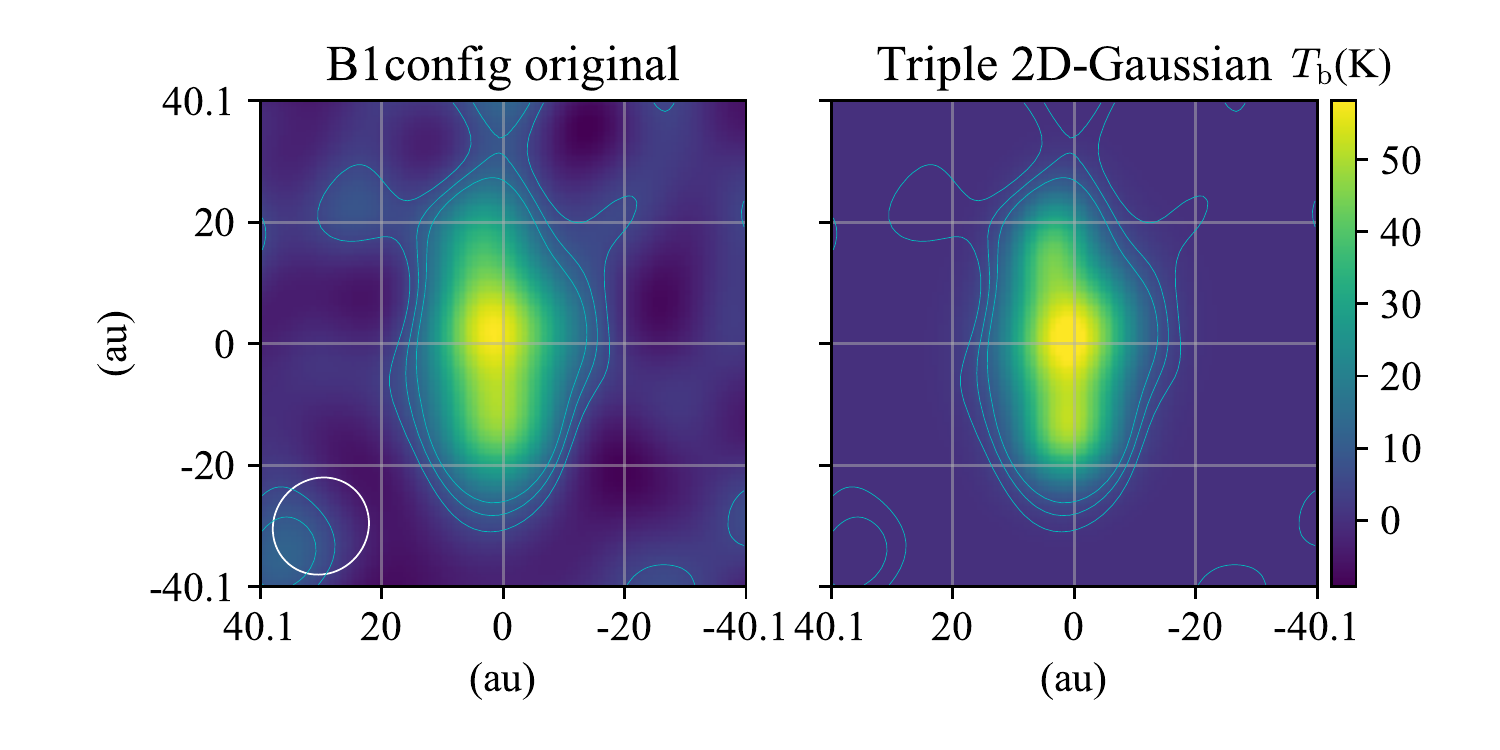}
        \includegraphics[width = \linewidth/2, clip]{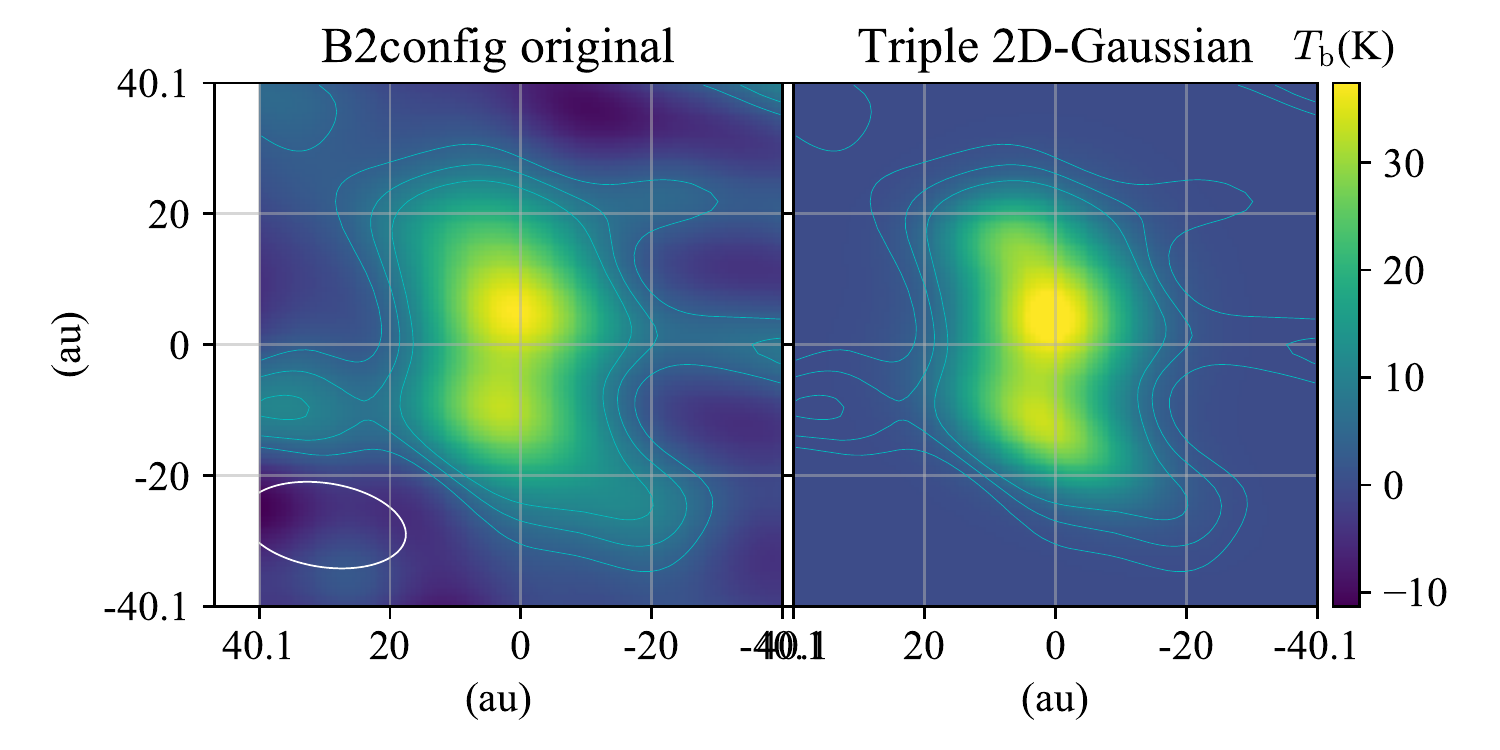}\includegraphics[width = \linewidth/2, clip]{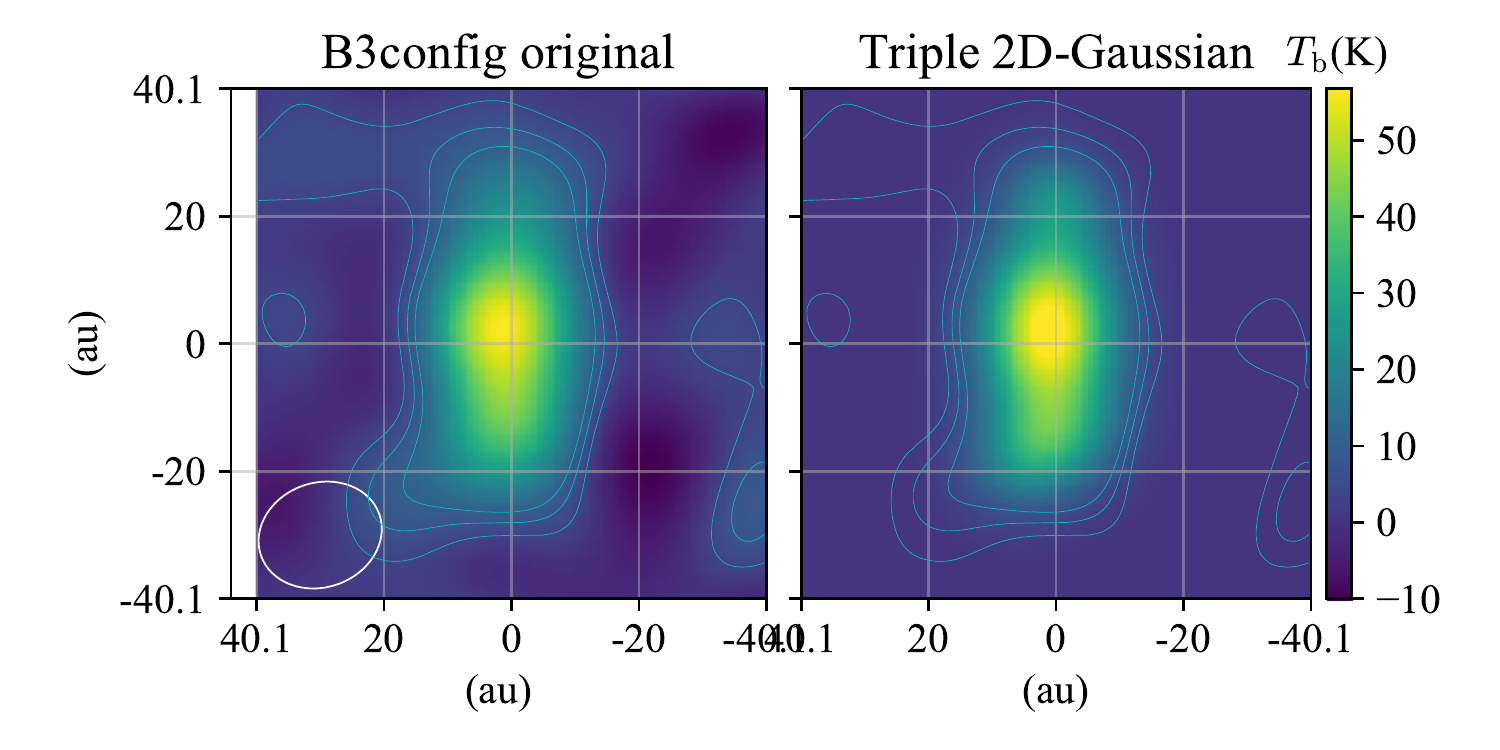}
        \caption{Original Q-band image and triple 2D Gaussian fit for each of array configurations. The cyan contours show 1, 2, 3$\sigma$ levels of the original images and are the same between the right and left panels.}
        \label{fig:2dgsfit}
    \end{figure*}
    
    \begin{table*}[htbp]
        \centering
        \begin{tabular}{ccCCCCCC}
             %&  \\
             %&
            Configuration& clump& a (\Kelvin) & x (\au) & y (\au) & \sigma_x (\au) & \sigma_y (\au) & {\rm P.A.\,(rad)}\\\hline \hline
            \input{fitsimaging_via_astropyThree2DGaussianFit}
        \end{tabular}
        \begin{tabular}{cCC}
             Configuration & D_{\rm NC} (\au) & D_{\rm CS} (\au)
            \\\hline 
            \input{fitsimaging_via_astropyThree2DGaussianFit_distances}
            \hline
        \end{tabular}        
        \caption{$D_{\rm NC}$ and $D_{\rm CS}$ are 
        the distances between clumps-N and -C,
        and between clumps-C and -S, 
        respectively, computed from the fitting results. }
        \label{tab:2dgsfit}
    \end{table*}
    
    \begin{figure*}
        \centering
        \includegraphics[width=\linewidth, clip]{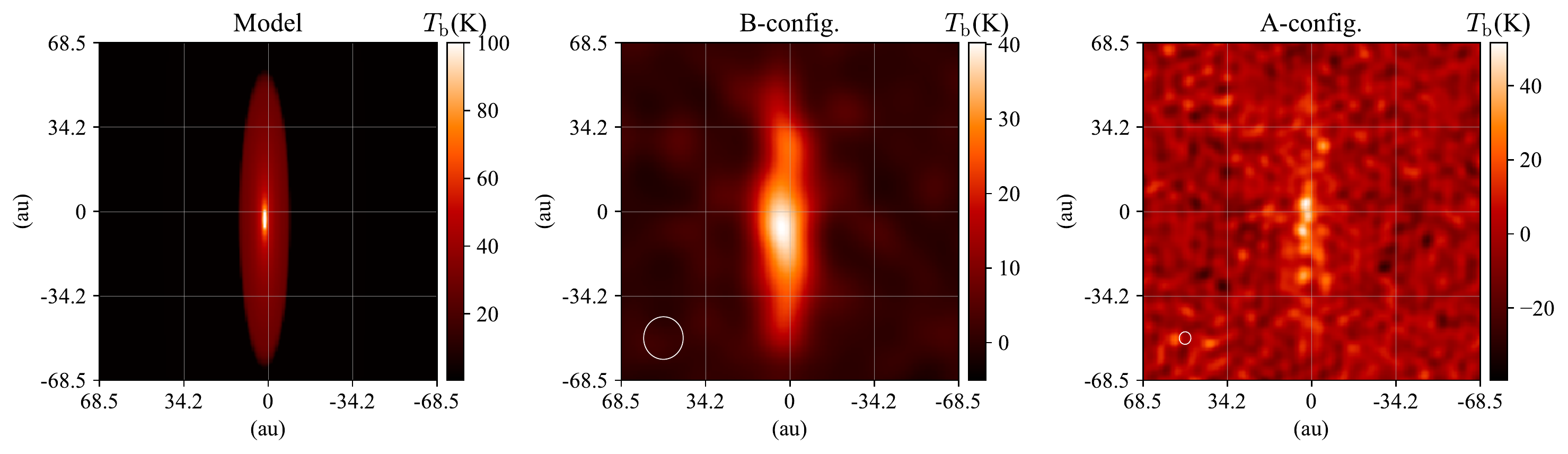}
        \caption{Examples of our synthetic observations. (left) model emission of a smooth disk with $\Sigma_0 = 10^4{\,\rm g}\cm^{-2}$ and $i = 10\deg$. The corresponding flux density is $\approx 4.9 {\rm \, mJy}$.
        (middle) observed image with VLA B-configuration and integration time of an hour. 
        Recovered flux density is $\approx 4.3 {\rm \, mJy}$
        (right) observed image with VLA A-configuration and integration time of two hours.
        Recovered flux density is $\approx 0.84 {\rm \,mJy}$.
        }
        \label{fig:simobserve}
    \end{figure*}
    
\end{document}

%% file: fitsimaging_via_astropyThree2DGaussianFit.tex
ABconfig& N &53.2 \pm 1.16&-0.596 \pm 0.0892&12.1 \pm 0.167&10.1 \pm 0.133&2.32 \pm 0.039&2.09 \pm 0.00429\\ 
& C &94.3 \pm 0.492&-2.22 \pm 0.0357&-1.39 \pm 0.0749&8.54 \pm 0.0952&4.65 \pm 0.0381&1.77 \pm 0.0108\\ 
& S &69 \pm 0.995&-1.19 \pm 0.0528&-17.3 \pm 0.0489&4.22 \pm 0.0616&2.43 \pm 0.0382&2.45 \pm 0.0163\\ 
\hline 
Aconfig& N &112 \pm 2.95&-1.5 \pm 0.0613&5.55 \pm 0.196&4.04 \pm 0.0966&5.16 \pm 0.0463&-1.95 \pm 0.0356\\ 
& C &120 \pm 3.21&-4.11 \pm 0.106&-2.27 \pm 0.137&3.88 \pm 0.0961&5.86 \pm 0.0443&-1.74 \pm 0.0208\\ 
& S &102 \pm 0.824&-2.93 \pm 0.038&-15.8 \pm 0.0521&4.4 \pm 0.0519&4.65 \pm 0.0377&-1.74 \pm 0.112\\ 
\hline 
B1config& N &32.3 \pm 0.527&-3.82 \pm 0.0445&16.6 \pm 0.118&6.38 \pm 0.0731&4.68 \pm 0.0532&1.29 \pm 0.0174\\ 
& C &59.6 \pm 0.247&-0.787 \pm 0.0271&0.671 \pm 0.0854&8.56 \pm 0.104&7.19 \pm 0.0291&1.27 \pm 0.0415\\ 
& S &39.6 \pm 0.618&-0.911 \pm 0.0433&-15.5 \pm 0.0951&5.89 \pm 0.0574&5.51 \pm 0.0465&0.901 \pm 0.0685\\ 
\hline 
B2config& N &16.2 \pm 0.487&-8.42 \pm 0.183&17.2 \pm 0.0913&4.77 \pm 0.105&7.24 \pm 0.127&1.67 \pm 0.0223\\ 
& C &39.5 \pm 0.158&-0.533 \pm 0.0532&3.96 \pm 0.111&9.1 \pm 0.109&8.67 \pm 0.0584&2.22 \pm 0.133\\ 
& S &27.3 \pm 0.317&-0.907 \pm 0.0855&-14 \pm 0.0819&5.15 \pm 0.046&10.8 \pm 0.0697&1.12 \pm 0.00619\\ 
\hline 
B3config& N &15.9 \pm 0.365&0.0758 \pm 0.0816&23.1 \pm 0.103&4.86 \pm 0.103&5.59 \pm 0.087&1.11 \pm 0.0705\\ 
& C &59.8 \pm 0.137&-1.09 \pm 0.0176&2.14 \pm 0.0814&7.17 \pm 0.0152&10.7 \pm 0.142&-0.0184 \pm 0.0059\\ 
& S &25.7 \pm 0.657&-2.05 \pm 0.0801&-16.5 \pm 0.101&7.63 \pm 0.0559&5.6 \pm 0.0759&0.441 \pm 0.0195\\ 
\hline 
Bconfig& N &27.1 \pm 0.685&-4.45 \pm 0.0563&15.3 \pm 0.234&8.98 \pm 0.0951&5.8 \pm 0.0584&1.2 \pm 0.0104\\ 
& C &51.6 \pm 0.475&-0.11 \pm 0.0639&1.07 \pm 0.134&9.66 \pm 0.0713&7.66 \pm 0.0669&0.974 \pm 0.0311\\ 
& S &33.4 \pm 0.541&-0.37 \pm 0.06&-15.2 \pm 0.107&6.91 \pm 0.0442&6.59 \pm 0.0556&2.74 \pm 0.083\\ 
\hline 

%% file: fitsimaging_via_astropyThree2DGaussianFit_distances.tex
ABconfig & 13.6 & 15.9 \\ 
Aconfig & 8.24 & 13.6 \\ 
B1config & 16.3 & 16.1 \\ 
B2config & 15.4 & 18 \\ 
B3config & 21 & 18.7 \\ 
Bconfig & 14.9 & 16.3 \\ 